\documentstyle[11pt,fleqn,aaspptwo,tighten]{article}
\newcommand{\sfRO}{\renewcommand{\baselinestretch}{1.1}\begin{small}}
\newcommand{\tfRO}{\end{small}\renewcommand{\baselinestretch}{1.6}}
\newcommand{\Sec}{\mbox{{\rm Sec. }}}

\newcommand{\etal}{\mbox{{\it et al. }}}
\newcommand{\negSMLskip}{\vspace*{ -3mm}}
\newcommand{\negMEDskip}{\vspace*{ -6mm}}

\newcommand{\negBIGskip}{\vspace*{-10mm}}
\newcommand{\disp}[2]{\mbox{$\sigma#2_{{\rm #1}}$}}

\newcommand{\Rc}{\mbox{$R_{\rm c}$}}

\newcommand{\abs}[1]{\mbox{$\mid #1 \mid$}}
\newcommand{\pmt}{\mbox{$\pm \;$}}

\newcommand{\rtp}[1]{\mbox{$^{#1}$}}

\newcommand{\Ht}{\mbox{${\rm H_2}$}}
\newcommand{\HI}{\mbox{{\rm H \footnotesize{I} }}}
\newcommand{\HII}{\mbox{{\rm H \footnotesize{II} }}}
\newcommand{\Msun}{\mbox{$M_{\odot}$}}
\newcommand{\Lsun}{\mbox{$L_{\odot}$}}
\newcommand{\hR}{\mbox{$h_{\rm R}$}}
\newcommand{\hRb}[1]{\mbox{$h_{\rm R #1}$}}
\newcommand{\ze}{\mbox{$z_{\rm e}$}}

\newcommand{\MSpcsq}{\mbox{$M_{\odot}{\rm pc}^{-2}$}}

\newcommand{\MoverL}[1]{\mbox{$M/{\cal L}_{{\rm #1}} \;$}}
\newcommand{\kms}{\mbox{${\rm km \;s}^{-1}$}}
\newcommand{\assq}{\mbox{${\rm arcsec}^2$}}
\newcommand{\passq}{\mbox{${\; \rm arcsec}^{-2} \; $}}
\newcommand{\asd}{\mbox{$''       \!\!.$}}
\newcommand{\amd}{\mbox{$'        \!\!.$}}
\newcommand{\add}{\mbox{$^{\rm o} \!\!.$}}
\newcommand{\as} {\mbox{$''       \;$}}
\newcommand{\am} {\mbox{$'        \;$}}
\newcommand{\ad} {\mbox{$^{\rm o} \;$}}
\newcommand{\tsd}{\mbox{$^{\rm s}  \!\!.$}}

\newcommand{\tm} {\mbox{$^{\rm m}\;$}}
\newcommand{\th} {\mbox{$^{\rm h}\;$}}
\newcommand{\VLAnote}{\footnote{The VLA of the 
National Radio Astronomy Observatory is a facility of the National
Science Foundation operated under cooperative agreement by Associated 
Universities, Inc.} \ }
\newcommand{\OvGnote}{\footnote{
Notice a typo in their Eq. (4): $\sigma_z(R)$ has to be replaced by
$\sigma_z(R) \times R$}
}
\newcommand{\SOTON}{Dept.  of Physics, University of Southampton, 
Southampton S17 1BJ, U.K.}
\chardef\Isp="10
\newcommand{\isp}{\mbox{\'{\Isp}}}

\chardef\isp="10

\begin{document}

\title{NGC 4244: A LOW MASS GALAXY WITH A FALLING ROTATION CURVE AND A
FLARING GAS LAYER}

\author{Rob P. Olling}
\affil{Columbia University, New York, \\
now at \SOTON \\
olling@astro.soton.ac.uk}

\begin{center}
\vspace*{2cm}
To be published in the Aug. 1996 issue of the Astronomical Journal
\end{center}

\authoraddr{\SOTON}

%%\clearpage

\begin{abstract}

I present sensitive high resolution VLA B, C, and D array observations
of the almost edge-on Scd galaxy NGC 4244 in the 21-cm spectral line of
neutral atomic hydrogen.  The gas layer of NGC 4244 is rather symmetric
in all respects, i.e.  the surface density distribution, flaring and
warping.  This symmetry allows for a reliable determination of the
rotation curve, despite the fact that the galaxy is close to edge-on. 
The rotation curve rises slowly in the inner 6 kpc, is roughly constant
at 100 \kms \ out to 10 kpc, and decreases in Keplerian fashion by 15\%
at the last measured point at 14 kpc.  The rotation curve constrains the
stellar mass-to-light ratio to lie between 50 and 100\% of the
``maximum-disk'' value.  

A new technique is presented to determine simultaneously the inclination
and the thickness of the gas layer from high resolution \HI
observations.  This procedure uses the apparent widths at many azimuths
(many channels) and can be used at inclinations as low as 60\ad. 
Kinematic information is used to separate flaring from warping.  The
inclination of the unwarped disk is about 84\add5, while the small warp
coincides with a decreasing inclination (to 82\add5 \pmt 1\ad).  The
data indicates that at large radii the disk warps back to the plane
defined by the inner disk.  The measured gaseous velocity dispersion is
roughly constant within the optical disk (8.5 \pmt 1 \kms) and increases
slightly beyond.

On both sides of the galaxy the thickness of the gas layer increases
gradually from $\sim$400 pc at 5 kpc to $\sim$1.5 kpc at the last
measured point (at 13 kpc).  The strong gradients in the {\em inferred}
thickness which bracket the spiral arms probably result from streaming
motions associated with the arms and are not intrinsic to the galaxy. 
In an accompanying paper (AJ, Aug. 1996) I use the measurements
presented in this paper to infer that the dark matter halo of NGC 4244
is highly flattened. 

\end{abstract}

%%%%\negMEDskip
\section{INTRODUCTION}
\label{sec-Introduction}
%%%%\negMEDskip

In a previous paper, Paper I (Olling 1995), it was shown that
measurements of the radial variation of the thickness of gas layers
(flaring) can be used to constrain the shape of the dark matter (DM)
halo.  This is accomplished by comparing the measured flaring with that
expected from a self-gravitating gaseous disk in a potential generated
by the stellar disk and (flattened) DM halo.  Here, high resolution \HI
observation of the nearby, almost edge-on, Scd galaxy NGC 4244 are
presented.  Using these observations I determine the rotation curve, the
flaring, the velocity dispersion, and the radial surface density
distribution of the gas.  In the accompanying paper (Paper III, AJ,
1996, ..., ...) I will apply the method presented in Paper I to
determine the flattening of the dark matter halo of NGC 4244 using the
essential parameters as derived in this paper.

In a study similar to the present one, Rupen (1991a) analyzed high
resolution \HI observations of two edge-on galaxies, NGC 4565 and NGC
891.  Due to the increased sensitivity of the VLA, sensitive high
resolution \HI observations, as pioneered by Rupen (1991b), can now be
routinely obtained and processed.  The high resolution \HI data
presented in this paper are sensitive enough to determine the gaseous
velocity dispersion and probe the region beyond the optical disk where
the thickness of the gas layer is most sensitive to the shape of the DM
halo.  The new technique to determine the thickness of the gas layer
(presented in \Sec \ref{sec-the.thickness.of.the.gas.layer}) complements
an existing procedure specific to the case of strong spiral structure in
the \HI (Braun 1991), and is more general than the method employed by
Irwin \& Seaquist (1991).

NGC 4244 is an intermediate mass system with a rotation speed of about
100 km/s.  Its proximity (3.6 Mpc) results in the high linear spatial
resolution (17.5 pc/arcsec) which is necessary to successfully determine
both the vertical structure of the gaseous disk and the rotation curve
in the inner parts of the galaxy.  The large inclination of NGC 4244
with respect to the line of sight, about 84.5 degrees, in combination
with the symmetric surface density distribution, allows for an accurate
determination of the rotation curve.  NGC 4244 is one of the few
galaxies to show, on both sides, the characteristic drop in rotation
velocity beyond the optical disk.

The outline of this paper is as follows: in \Sec
\ref{sec-the.galaxy.NGC4244} I summarize current observational knowledge
of NGC 4244.  The \HI observations are presented in \Sec
\ref{sec-HI.observations} and the parameters derived therefrom in \Sec
\ref{sec-Inferred.properties}.  A new technique to derive the thickness
and inclination of the gas layer is outlined in \Sec
\ref{sec-Outline.of.the.method}, and is applied in \Sec
\ref{sec-thickness.of.NGC4244s.gas.layer}.  The results are discussed in
\Sec \ref{sec-Discussion}.

%\subsection{The galaxy NGC 4244}
%\input{gala_02}
%
% File name : gala_02.tex
%
%%%%\negMEDskip
\section{THE GALAXY NGC 4244}
\label{sec-the.galaxy.NGC4244}
%%%%\negMEDskip

NGC 4244 is a small almost edge-on ScdIII galaxy in which individual
stars and \HII regions can be resolved (Sandage \& Bedke 1994).  Surface
brightness photometry (van der Kruit \& Searle 1981a) reveals that the
radial light distribution is close to exponential till about five
optical scale lengths, where it cuts off sharply (see also Fig. 
\ref{fig:radial.light.profile}).  The vertical scale height of the light
is constant with radius, while no thick disk is present.  The very faint
bulge ($\mu_{\rm K}(r_{\rm e}) = 23.05$ mag/\assq) with an effective
radius of 140\as (Bergstrom \etal 1992) is dynamically insignificant.

There are several indications that the current star formation rate is
very low in NGC 4244.  Only very low radio continuum emission (Condon
1987) and little H$\alpha$ emission (Walterbos 1994, private
communications) have been detected, while upper limits have been set on
X-ray emission (Bregman \& Glassgold 1982) and emission from highly
ionized halo gas (Deharveng \etal 1986).  Very faint radio continuum
emission extends along the plane of the galaxy from 2\amd5 north-east
(NE) to 2\am south-west (SW) of the center.  The 60$\mu$m emission is
enhanced in this region as well (Rice 1993).  In contrast to most
galaxies (Bicay \& Helou 1990), NGC 4244's 20 cm radio continuum
emission is not more extended than the 60$\mu$m emission.  The weak CO
emission (Sage 1993), coincident with the radio continuum emission,
amounts to 2.2 x 10$^7$ \Msun \ of \Ht \ (50\% of the \HI mass in this
region).  The \Ht \ radial surface density distribution is calculated
with the Abel inversion technique (\Sec \ref{sec-Abel.inversion}) and is
consistent with a Gaussian ring centered at 1.3 kpc, with a FWHM of 1.6
kpc and an amplitude of 1.6 \MSpcsq.

Throughout this paper the distance to NGC 4244 is taken to be 3.6 Mpc,
consistent with the IR-Tully-Fisher distance (3.7 Mpc, Aaronson \etal
1986) and the ``hydrogen counterpart'' of the Tully-Fisher distance (3.6
Mpc) as defined by Broeils (1992, Chapter 3).  NGC 4244 is the second
brightest member of a group of dwarf galaxies: the CVn I (de Vaucouleurs
1975) or B4 (Kraan-Korteweg \& Tammann 1979) group with 21 catalogued
members.  This group has a radius of 1 Mpc, an average density of 4.5
Mpc$^{-3}$, and a total luminosity of $8.9 \times 10^9$ \Lsun$_{,B}$. 
The nearest catalogued neighbor is the SmIV galaxy NGC 4190 at a
projected distance of 81 kpc.  This group is rather elongated (axial
ratio $\sim$ 3), with the position angle of the long axis at -18\ad,
66\ad \ from the plane of NGC 4244.  The observation presented in this
paper (\Sec \ref{sec-HI.distribution}) fail to detect gas rich
companions more massive than 10$^7$ \Msun \ within 45 kpc.

%\subsubsection{Optical Data}
\label{sec-optical.data}

Figure \ref{fig:IIIaJ.image} shows the optical (grey scale) image from
IIIaJ photographic observations (van der Kruit \& Searle 1981a) and a
total hydrogen map.  The orientation of the weak dustlane suggests that
the eastern side of the galaxy is the far side and thus least affected
by absorption.  The intrinsic radial surface brightness distribution,
$L(R)$, is derived by applying the Abel-inversion technique (\Sec
\ref{sec-HI.surface.density.distribution}) to four different strips
parallel to the major axis, after removal of foreground stars.  The
first strip contains all light perpendicular to the minor axis (top
profile in Fig.  \ref{fig:radial.light.profile}).  The second strip
contains all light within 420 pc (twice the vertical exponential scale
height, \ze) from the midplane.  The third and forth strips are
determined from those parts which are further than 420 pc from the
midplane (east and west respectively).  As the region which is affected
by dust extinction lies generally close to the plane ($z \la \onehalf
\ze$ ;Peletier \etal 1995), the third and forth strips are expected to
be free of extinction effects.  All four profiles indicate a radially
exponential light distribution with the same radial scale length (\hR)
of 2.0 kpc, slightly larger than the value derived by van der Kruit \&
Searle (1981a)\footnote{All the linear distances derived by van der
Kruit \& Searle (1981a), who use a distance of 5 Mpc (Sandage \&
Tammann, 1975), are scaled to the adopted distance for NGC 4244, 3.6
Mpc.}.  Thus, the overall light distribution is probably not much
affected by extinction (the adopted fit is indicated by the filled
circles).  As noticed by van der Kruit \& Searle (1981a), the stellar
disk is truncated at somewhat different radial distances on the two
sides of the galaxy.  The IIIaJ-magnitudes are converted to B-magnitudes
assuming a (B-V)=0.7 as appropriate for the old disk (Bottema 1989). 
Some parameters of this galaxy are listed in Table
\ref{tab:Table.N4244.parameters}. 

%--------------------     Table I goes here  --------------------

%\subsection{HI Observations}
%\input{obst_02}
%
%  File: obst_02.tex
%
%%%%\negMEDskip
\section{\protect\HI OBSERVATIONS}
\label{sec-HI.observations}
%%%%\negMEDskip

NGC 4244 was observed with the VLA\VLAnote \\ (Napier \etal 1983), in the
10 km (B), 3 km (C), and 1 km (D) arrays for a total of about 14 hours. 
The data from 1989 and 1990 were taken as part of a project aimed at
doing a multi-wavelength high resolution study of the interstellar
medium in nearby galaxies (Braun 1995).  The observing parameters are
summarized in Table \ref{tab:Table.N4244.obs.parameters}. 

  Standard calibration procedures were followed.  After removal of some
data affected by inter- \\
ference\footnote{In particular the 1992 D-array
daytime observations required extensive UV-editing due to solar
interference.} the data were combined in the UV plane.  The primary flux
calibrator was 3C 286, with an assumed flux at 1418 MHz of 14.86 Jy.  A
heavily tapered continuum image was made using 5 line-free channels on
both sides of the band.  The rms noise in this image is 0.94 mJy
beam\rtp{-1} at a resolution of 80\as x 77\as.  Several background
sources but no extended radio continuum emission was detected from NGC
4244 itself down to a 3 sigma limit of 2.8 mJy beam\rtp{-1}, consistent
with a peak flux of 1.4 mJy as determined by Condon (1987) and Hummel
\etal's (1984) non-detection at 610 MHz.  A strong background source
about 7\arcmin \ NE of NGC 4244 is probably the reason for the much
stronger detection reported by Schlickeiser \etal (1984). 

% --------------------     Table II goes here     --------------------

For the two high resolution data sets the sensitivities listed in Table
\ref{tab:Table.N4244.obs.parameters} correspond to roughly $3 \times
10^4$ \Msun \ beam\rtp{-1} channel\rtp{-1} at the field center, while the
two low-resolution sets are about three times more sensitive as a result
of the interim upgrade of the VLA receivers.  To bring out emission at
different scales, images of the line emission were made with different
tapering.

\subsection{Results}
\label{sec-HI.distribution}
\label{sec-The.Global.Profile}

Assuming low optical depth the total amount of neutral hydrogen is
arrived at by adding the flux containing areas in the channel maps.  The
validity of this assumption is checked in Appendix
\ref{sec-Optical.Depth.Effects}.  The flux containing areas are defined
to be those regions which have a flux larger than $\sim 1.2 \sigma$ in a
smoothed version of the spectral line cube (smoothed to a four and three
times lower resolution, spatially and in velocity respectively).  The
high and low-resolution total \HI distributions are presented in Fig. 
\ref{fig:NGC4244.total.HI}.  The images have been rotated 48\ad \
clockwise, so that the upper part is the north-eastern side.  Both the
high and low resolution data show a warp starting at the edge of the
optical disk (about 9 arcmin from the center).  The \HI distribution of
NGC 4244 is rather symmetric apart from the ``popped blister'' located
just below the major axis about four arcmin to the north-east.  A
recent star formation event might be responsible for this feature as it
coincides with peaks in the radio continuum and IRAS emission. 

The low resolution, primary beam corrected, total intensity image yields
a total \HI mass of $(1.29 \pm 0.05) \times 10^9$ \Msun.  The
north-eastern part contains 51\% of the total \HI mass, the
south-western part amounts to 49\%.  The hydrogen mass determined from
the global profile (the flux as a function of velocity, Fig. 
\ref{fig:global.profile}), $(1.4 \pm 0.1) \times 10^9$ \Msun, is
consistent with the determination from the total \HI map.  The \HI mass
listed in Table \ref{tab:Table.N4244.HI.parameters} is the average of
the total \HI and global profile values, where the error equals half the
difference.

From the global profile a systemic velocity (the mean of the 10\%, 20\%
and 50\% points) of 244.4 \pmt 0.3 \kms \ is determined, in agreement with
the value determined by Huchtmeier \& Seiradakis (1985).  As a result of
the asymmetry in the \HI distribution, the receding part has a higher
peak value than the approaching part.  The position at the major axis
where the recession velocity equals the systemic velocity is offset by
$7\asd4 \pm 0\asd6$ towards the south-west, which agrees well with the
central position of the \HI as determined in \Sec
\ref{sec-HI.surface.density.distribution}.

Within 45 kpc from the pointing center, no companions were detected down
to a five sigma level of 10$^5$ \Msun \ channel\rtp{-1} at the center
(or $1.6 \times 10^7$ \Msun \ channel\rtp{-1} at 45 kpc).

% --------------------     Table III goes here     --------------------

%\subsubsection{The Channel Maps}
\label{sec-The.Channel.maps}

Fig.  \ref{fig:channel.maps} shows a composite of the high resolution
(28\asd3x10\asd1 or 490x177 pc along the major and minor axes
respectively) and intermediate resolution (38\as $\!$x38\as $\!$)
channel maps.  The maps are combined such that the channels with the same
speed relative to the systemic velocity are present in one frame (after
having blanked those areas without emission).  The receding part of NGC
4244 is the south-western part (negative X-values), the north-eastern
part is approaching.  The low velocity channels, which would overlap if
presented in this manner, are presented individually.  One local
irregularity is clearly visible: the +4\arcmin \ ``hole'' found in the
total \HI map can be seen in the frames labeled 17,18, and 19. 

Each map shows a characteristic ``filled V'' shape.  Two different
mechanisms can cause such shapes.  First of all it reflects the flaring
of the gaseous disk.  The other contributing factor is kinematic in
nature: the ``butterfly'' pattern (seen in channel maps of less inclined
systems, e.g., Bosma 1981) seen almost edge-on.  In \Sec
\ref{sec-Outline.of.the.method} I present a new technique to disentangle
these two effects.  Classically the flaring is measured in the edge
channels, i.e.  at velocities of $(V_{\rm sys} \pm V_{\rm rot,max})$
(Sancisi \& Allen 1979; van der Kruit 1981; Rupen 1991a).  Because NGC
4244 is not exactly edge-on and because the velocity edge depends on
radius (due to the falling rotation curve), the apparent widths can not
be easily interpreted in terms of an intrinsic vertical scale height. 
Furthermore, systematic errors, e.g.  due to non-circular motions are
largest in the edge channels (Appendix
\ref{sec-The.effects.of.kinematical.distortions}).  The channels $\pm$18
are the edge channels between about 5\am and 9\am.  For this radial
range, the receding side exhibits a more-or-less regular increase of
apparent width, but on the approaching side this effect is not evident. 
This difference can easily arise from local effects such as arm
inter-arm surface density differences, holes, bubbles, and streaming
motions.  The new thickness-and-inclination determination technique
outlined in \Sec \ref{sec-Outline.of.the.method} minimizes the effects
of such irregularities.

%\subsubsection{The Warp}
\label{sec-The.Warp}

A rather regular warp, starting just before the optical edge of the
galaxy, is seen in the channel maps.  The channels $\pm$16,$\pm$17, and
$\pm$18 show the strongest warping at large radii.  The position angle
of the warp is determined by tracing the major axis of emission as a
function of major axis distance.  Within the optical disk, the position
angle of the centroid is constant at 48\ad east of north.  At a radius
of 8\am, the position angle begins to change linearly to reach 43\ad at
13\am.  Because no clear butterfly pattern is seen in the warped region,
the warp is likely to be the result of a change in position angle rather
than in inclination.  This is confirmed by the derived radial variation
of the inclination (Fig.  \ref{fig:Thickness.and.Inclination}).

%\subsection{Extractions}
%\input{extra_02}
%
%  File: extra_02.tex
%
%%%%\negMEDskip
%\subsection{INFERRED PROPERTIES}
\section{INFERRED PROPERTIES}
\label{sec-Inferred.properties}
%%%%\negMEDskip

In this section several properties of the gas\footnote{Unless otherwise
noted, the term ``gas'' refers to \HI in an observational context, and
to the sum of \HI, \Ht \ and He in a dynamical context.} layer will be
determined: the radial surface density distribution ($\Sigma_{\rm
gas}(R)$), the rotation curve ($V_{\rm rot}(R)$), and the radial
variation of the gaseous velocity dispersion ($\disp{gas}{}(R)$).  All
these quantities are required to make self-consistent predictions of the
radial variation of the thickness of the gas layer ( Paper I, Paper III). 
Furthermore, with the aim to gauge the reliability of the determination
of these properties, these quantities are used to compare the VLA
observations with simulated observations of NGC 4244 (Appendix
\ref{sec-VLA.simulate}). 
 
\label{sec-The.center.of.mass}

Integrating the total hydrogen along the minor axis results in a major
axis \HI profile ($\Lambda(x)$).  The inner 4\am of this profile are
symmetric about a point located $6\asd9 \pm 0\asd6$ to the south-west
(along the major axis) of the catalogued position.  Similarly, the
location of the midplane is offset by $2\asd2 \pm 0\asd5$ to the
south-east.  In the rest of this paper I assume that NGC 4244's center
is located at the centroid of the \HI-distribution: $\alpha$ = 12\th
15\tm 9\tsd7, $\delta$ = 38\ad 4\am 59\asd4, which position is listed in
Table \ref{tab:Table.N4244.HI.parameters}.  This position agrees with
the kinematic center determined in \Sec \ref{sec-The.Global.Profile}.

%%%%\negSMLskip
\subsection{The \protect\HI Surface Density Distribution}
\label{sec-HI.surface.density.distribution}
\label{sec-Abel.inversion}
\label{sec-coordinate.system}
%%%%\negSMLskip

I describe the line of sight deconvolution necessary to determine the
radial surface density distribution after introducing the coordinate
systems used.  First there is the coordinate system centered on the
center of the galaxy: ($x, y, z$) in Cartesian coordinates and
($R,\theta,z$) in polar coordinates (with $\theta=0$ being the
intersection of the plane of the sky with the plane of the galaxy, and
with $z$ the vertical coordinate).  In the observers coordinate frame,
$x$ and $d$ are coincident with the major and minor axes of the inner,
unwarped, part of the disk.  The projection of the line of sight onto
the (unwarped part of the) plane of the galaxy is parallel to the
$y$-axis. 

I determine an approximation to the \HI profile (the total \HI
distribution summed perpendicular to the major axis) by summing the
emission over the $d$-axis.  Because of the warp the $x$-coordinate does
not truly trace the major axis ($x'$).  However, since $x' = \sqrt{x^2 +
d^2}$ does not differ significantly for the small warp of NGC 4244 ($d
\leq$ 1 kpc for $x \leq 13$ kpc), the derived density distribution is
not affected.  The integration\footnote{Assuming low optical depth, see
Appendix \ref{sec-Optical.Depth.Effects}.} of the volume density along
the line of sight and perpendicular to the major axis, which yields the
\HI profile, can be written as an integration of the true surface
density along the $y$-axis:

\negMEDskip
\begin{eqnarray}
&& \hspace*{-12mm} 
   \Lambda(x') \; \approx \;
   \int_{-\infty}^{+\infty}  \; dy \; \Sigma_{\rm gas}
       \left( \sqrt{x^2+y^2} \; \right)  \nonumber \\
&& \hspace*{-7mm} 
=  \; 2 \times 
   \int_{x}^{\infty}
       \frac{R \; \Sigma_{\rm gas}(R)}{\sqrt{R^2-x^2}} \; dR \hspace{3mm}
       ( \rm M_{\odot}{\rm / pc} ) \; .
\end{eqnarray}
\negMEDskip

\noindent Inversion of this Abel integral [e.g., Binney \& Tremaine,
Eq.  (1B-59b)] yields the radial surface density distribution

\negMEDskip
\begin{eqnarray}
&& \hspace*{-1cm}
   \Sigma_{\rm gas}(R) \; = \; -\frac{1}{\pi} \times  \nonumber \\
&& \hspace*{-1cm}
   \int_{R}^{\infty} 
   \frac{1}{\sqrt{\left( x^2 - R^2 \right)}} \;
   \frac{d\Lambda(x)}{dx} \; dx \hspace{5mm}
       ( \rm M_{\odot}{\rm / pc^2} ) \; .
   \label{eq:Sigma.gas.R}
\end{eqnarray}
\negMEDskip

\noindent This procedure can be applied at any wavelength for any
inclination provided that the optical depth is low, the galaxy is not
warped and if it is axisymmetric.  Note that similar techniques were
used to determine the surface brightness distribution of Virgo spirals
from ``pencil-beam'' \HI data (Warmels 1988) and the mass-to-light ratio
of elliptical galaxies (i.e., Schwarzschild 1954; Binney \& Mamon 1982). 
Adding realistic noise to a simulated spectral line cube (Appendix
\ref{sec-VLA.simulate}) and applying Eq.  (\ref{eq:Sigma.gas.R}) yields
the input $\Sigma_{\rm gas}(R)$.  Thus, the derivation of the surface
density distribution is not sensitive to noise.  However, if the signal
to noise ratio is poor, it may be preferable to use the iterative
technique proposed by Warmels (1988). 

$\Sigma_{\rm gas}(R)$ is derived from a combination of the high and low
resolution $\Lambda(x)$ curves and is presented in Fig. 
\ref{fig:Sigma.gas.R}.  Negative surface densities at small radii occur
unless the two sides of the $\Lambda(x)$ profile are averaged over the
inner $\sim$ 1.5 kpc.  Thus this method does not recover the true
$\Sigma_{\rm gas}(R,\theta)$ but it can find an azimuthally symmetric
density distribution consistent with the observations: it serves as a
reasonable approximation to the true distribution.  Integrating the
$\Sigma(R)$ profiles out to 15.8 kpc (where the surface density equals
0.02 \MSpcsq), I find $1.4 \times 10^9$ \Msun: the Abel inversion
technique conserves flux.

Several strong (up to a factor 3) density enhancements ($\sim$1 kpc
wide), possibly spiral arms, are present.  In the inner 4 kpc the
south-western and north-eastern density distributions are very similar. 
Beyond 10 kpc the surface density decreases in an exponential manner
with a scale length of roughly 1.36 kpc.  No sharp cutoff in the
\protect\HI column density is seen {\em above} the 3$\sigma$ detection
limit ($5 \times 10^{18}$ cm$^{-2}$).  We see that the surface density
distribution is rather symmetric (see also \Sec
\ref{sec-thickness.of.NGC4244s.gas.layer}), except in the 6 - 8 kpc
region.  It might be that this surface density difference causes
differences in the observed flaring (Fig. 
\ref{fig:Thickness.and.Inclination}) as a result of the different
self-gravity at these points. 

%%%%\negSMLskip
\subsection{The Rotation Curve and the Velocity Dispersion of the Gas}
\label{sec-rotcur.and.velocity.dispersion}
%%%%\negSMLskip

Figure  \ref{fig:XV-plots} shows the intensity as a function of velocity
and position along the major axis (summed over the minor axis).  The
rotation curve and velocity dispersion of NGC 4244 are derived from this
plot.  The maximum relative velocity outline corresponds roughly to the
rotation speed minus twice the velocity dispersion in the tangential
direction.  Approximations to the rotation speed and velocity dispersion
were determined by fitting a single Gaussian to the extreme velocity
envelope.  These approximations have been corrected for instrumental,
line-of-sight integration, and beam smearing effects, as detailed in
Appendix \ref{sec-Theoretical.spectra}. 

Figure \ref{fig:gaseous.velocity.dispersion} shows NGC 4244's smearing
corrected intrinsic tangential dispersion as derived from the high,
intermediate and low resolution data sets, which sample different ranges
in galactocentric radius.  For each set of measurements, both sides
yield dispersions generally equal to within the errors.  The errors on
the values extracted from the 10\asd \ data set are large because the
signal to noise ratio is relatively low.  The average dispersion curve
is the weighted average of the measurements obtained at the two sides
(38\as \ data set for $R \in (9,13]$ kpc, and 55\as \ beyond).  For
radii smaller than 9 kpc I also average the measurements from the
Hanning-smoothed 10\asd1 and 38\as \ data sets since these data arise
from essentially independent measurements (B \& D array, respectively).

In Fig.  \ref{fig:Rotation.curves} I present the line-of-sight and
beam-smearing corrected rotation curves for both sides of the galaxy. 
Apart from the region between 2.5 and 7.5 kpc, the rotation curves are
equal to within the errors.  The drop in the north-eastern rotation
curve seen between 2.5 and 5.5 kpc is clearly a result of a local
phenomenon (described above): there the adopted rotation curve follows
the south-western curve (10\as x10\as) till 4.5 kpc.  The adopted
rotation curve between 4.5 and 6 kpc is a smooth junction of the
adjacent regions: here the rotation speeds as determined from the XV
plots are probably influenced by the strong \HI peak at 5.5 kpc (see
Appendix \ref{sec-Theoretical.spectra} for a discussion).  The 6 to 7.5
kpc region, on the south-western side, corresponds to the region where
the B-band surface brightness is depressed by a factor of about 10. 
Therefore, I choose the intrinsic rotation curve to follow the
north-eastern curve in this range of galactocentric radii (38\as
x38\as).  Taking these modifications into account leads to the adopted
intrinsic rotation curve (indicated by the filled circles in Figure
\ref{fig:Rotation.curves}), which is minimally affected by non-circular
motions. 

Because of the large inclination, small changes thereof are not
important for the derivation of the rotation curve.  The maximum
amplitude of the rotation velocity is about 100 \kms \ and starts
dropping significantly beyond $R_{\rm max}$, the edge of the optical
disk.  At a distance of 14 kpc the rotation speed has dropped to 85
\kms.  Huchtmeier (1975) finds, using a 9\am beam, that the \HI extends
out to 18\am and determines a rotation speed of $85 \pm 5$ \kms \ at
that distance.  The measurements presented in this paper do not confirm
this detection.  The symmetry between the two sides indicates that the
gas is likely to be on circular orbits, except for the regions mentioned
above.  The symmetric drop in rotation curve on both sides of the disk
is the strongest argument that we are actually measuring the circular
velocity and that this drop is not caused by, for example, a lack of gas
on the major axis: an explanation which has been invoked to explain
asymmetric drops seen in other edge-on galaxies (Sancisi \& Allen 1979).

%\subsection{The thickness of the HI layer}
%\input{thick_02}
%
% File: thick_02.tex
%
%%%%\negMEDskip
\section{THE DETERMINATION OF THE \\
THICKNESS OF THE HYDRO- \\ 
GEN LAYER}
\label{sec-the.thickness.of.the.gas.layer}
%%%%\negMEDskip

In the past, the radial increase of the thickness of the gas layer of
external galaxies has been determined from apparent width measurements
of the edge channels only (Sancisi \& Allen 1979; van der Kruit 1981;
Rupen 1991a).  The main advantage of this procedure is that the edge
channels select a region (the ``emitting region'') around the major axis
which is relatively narrow, so that its width, as projected onto the
sky, does not contribute much to the observed apparent thickness.  This
procedure has two disadvantages however: 1) the magnitude of the
projection effect depends strongly upon inclination, e.g., at i=85\ad \
it varies by a factor of two for an inclination difference of 2\ad, and
2) only two points per galactocentric radius, one on the approaching
side and one on the receding side, are measured. 

Irwin \& Seaquist (1991) modeled the entire spectral line cube obtained
for NGC 3079 to obtain the rotation curve, surface density distribution
and gas layer width (assumed to be constant with radius).  Braun (1991)
devised a technique to determine the inclination and the thickness of
the gas layer for M31, where the \HI emission is found primarily in the
spiral arms.  The technique presented here works best if the \HI is
distributed smoothly. 

Below I describe a new technique to simultaneously determine the radial
variation of the inclination and the thickness of the gas layer of
external galaxies.  As the essence of this method is to utilize data
from many azimuths (channels), both disadvantages of the edge-channel
procedure are circumvented.  This is accomplished by modeling the
contribution of the emitting region to the apparent width.  Two aspects
are crucial: 1) the emitting region can be well modelled if the rotation
curve is known, for example for a flat rotation curve a channel of a
several \kms \ wide is sensitive to gas in a wedge-like region in the
plane of the galaxy (Sancisi \& Allen 1979, their Fig.  6), where the
width of this emitting region follows directly from the rotation curve,
the observing velocity and velocity resolution, and 2) for a given
galactocentric radius the planar width of the emitting region varies
systematically with azimuth (i.e., observing frequency, or velocity). 
Thus the projected width of the emitting region also varies
systematically with azimuth and radius in a manner which can be
described exactly for a given observed rotation curve and channel width. 
These projection effects can be clearly seen in Fig.  8 of Sancisi \&
Allen, where the apparent width of the gas layer is larger in those
channels whose velocities are closer to the systemic velocity.  The
apparent width of the gas layer is then the sum of the projected
vertical thickness of the gas layer and the projected planar width of
the emitting region.  Only two unknowns (the vertical thickness and the
inclination) are involved.  Since the apparent widths can be measured at
many azimuths, and the planar width of the emitting region can be
calculated as a function of galactocentric radius and azimuth, the
inclination and vertical thickness of the gas layer can be
simultaneously determined from measurements of the apparent thickness in
many channels.

%%%%\negSMLskip
\subsection{Outline of the Procedure}
\label{sec-Outline.of.the.method}
%%%%\negSMLskip

The procedure outlined below works for a more realistic situation where the
rotation curve is not flat and where the gaseous velocity dispersion
softens the edges of the wedges.  A more detailed description is given
elsewhere (Olling 1995b), and is applied in the next subsection.

In the case where a galaxy is not edge-on, the azimuthal dependence of
the flaring might be determined from measurements of the apparent
thickness (in a given velocity channel) along an ellipse with axial
ratio $1/\cos{i}$.  For a close to edge-on galaxy this procedure is
difficult, while at $i=90\ad$ it fails altogether.  However, at any
inclination, the galactocentric radius ($R$) can be determined from the
rotation curve ($V_{\rm rot}(R)$), the major axis position ($x$) and the
observing velocity ($V_{\rm chan}$):

\negMEDskip
\begin{eqnarray}
R &=& \frac{x}{\cos{\theta}}
   = x \; \frac{V_{\rm rot}(R)\; \sin{i(R)}}
               {\abs{ V_{\rm chan} - V_{\rm sys}}}
   \nonumber \\
  &=& x \; \frac{V_{\rm proj}(R)}
               {\abs{ V_{\rm chan} - V_{\rm sys}}}
   \;, \label{eq:Rproj}
\end{eqnarray}
\negSMLskip

\noindent with $i(R)$ the inclination, and $V_{\rm sys}$ the galaxy's
systemic velocity.  Here the assumption is made that the galaxy can be
described adequately by a tilted ring model (Rogstad \etal 1974; Begeman
1989) which implies that the warp angle ($i(R)$) does not depend on
azimuth.  It is important to realize that the determination of $R$
depends on the {\em projected} rotation speed (i.e.  the {\em product}
of $V_{\rm rot}(R)$ and $\sin{i(R)}$) which is, at any inclination, much
better determined than the individual quantities (Begeman 1989): the
kinematic determination of $R$ is robust\footnote{The effects of
non-circular motions are discussed in Appendix
\ref{sec-The.effects.of.kinematical.distortions}.}.  When the apparent
widths in a given channel are determined, Eq.  (\ref{eq:Rproj}) is used
to assign a galactocentric radius to the measurements obtained at major
axis position $x$.  For a given channel and major axis position, the
planar width of the emitting region can be determined as well, so that
the apparent width measurements can be corrected for inclination and
projection effects to yield the intrinsic thickness of the gas layer. 
Below I describe in detail how the width of the planar emitting region
as well as the inclination and thickness of the gas layer can be
determined. 

Rather than using the wedge-approximation discussed in previous
sections, I include the effects of the velocity dispersion of the gas,
the exact shape of the rotation curve, and the gaseous surface density
distribution to determine the planar width of the emitting region. 
Given these parameters it is easy to calculate, for all points in the
galaxy ($x,y$), what fraction of the total column density ($\Sigma(R)$)
can be detected when observing the galaxy at inclination $i(R)$ and
velocity\footnote{The Channel Response Function appropriate for these
VLA observations is used (see Appendix
\protect\ref{sec-Theoretical.spectra}).} $V_{\rm chan}$ (Olling 1995b,
Chapt.  4).  These channel surface density distributions (CSDDs) form a
three dimensional data set ($\Sigma(x,y,V_{\rm chan})$), where each
plane represents the face-on surface density distribution which has
(given the galaxian properties and the observing geometry) the right
radial velocity to be observed in the channel centered on $V_{\rm
chan}$.  If observed face-on for example, the emission observed in each
channel is a fraction (which depends on the gaseous velocity dispersion
and the channel response function, but is constant across the galaxy) of
the surface density.  Some CSDD maps are presented in Fig. 
\ref{fig:CSDDs}.  Alternatively, one could think of a CSDD as a
deprojected (to face-on view) channel map of an infinitely thin gas
layer.  A rough idea about the shape of the CSDDs can be formed by
deprojecting (to face-on view) a velocity field which is contoured at
$MAX(\Delta V_{\rm chan},\disp{gas}{})$ \kms \ intervals and selecting
the area between two contours (with $\Delta V_{\rm chan}$ the width of
the channels).  But note that a CSDD is neither a deprojected channel
map nor a deprojected velocity field.  Along a line perpendicular to the
major axis, the shape of the emitting region is no longer discontinuous,
but can be parameterized, to reasonable accuracy, by two Gaussians:

\negMEDskip
\begin{eqnarray}
&& \hspace*{-12mm} 
   \Sigma_{\rm C}(y;x,V_{\rm chan}) \; \; \propto \; \; 
   \left( 
      e^{-\frac{1}{2}((y-Y_{\rm C})/W_{\rm C})^2} \; + \right. \nonumber \\
&& \hspace*{5mm} \left.
      e^{-\frac{1}{2}((y+Y_{\rm C})/W_{\rm C})^2}
   \right)
   \; , \label{eq:Equation.For.Sigma.C}
\end{eqnarray}
\negMEDskip

\noindent one for the near side and one for the far side.  Non-linear
least squares fits to the profiles extracted from Fig.  \ref{fig:CSDDs}
yield typical errors of 0.5\% - 1.5\% and 0.2\% - 0.6\% for $W_{\rm C}$
and $Y_{\rm C}$, respectively.  As the CSDDs are different for each
channel, so are (for a given galactocentric radius) $Y_{\rm C}$ and
$W_{\rm C}$, which depend mainly on the rotation curve and the gaseous
velocity dispersion, respectively.  Projecting the CSDD onto the sky,
the apparent planar width is given by:

\negMEDskip
\begin{eqnarray}
&& \hspace*{-12mm}
   \cos{i} \; FWHM_{\rm p}(x;V_{\rm chan}) \; \approx \; 2.35 \cos{i} \; \times
   \nonumber \\
&& \hspace*{-2mm}
   \sqrt{ W_{\rm C}^2(x;V_{\rm chan}) + \left[ Y_{\rm C}(x;V_{\rm chan}) 
   \right]^2}
   \; , \label{eq:Equation.For.FWHMp}
\end{eqnarray}
\negMEDskip

\noindent where the $Y_{\rm C}$-term is only included when the
projections of the near and far side of the CSDD are unresolved by the
beam, i.e.  in nearly edge-on viewing geometry.  Since the vertical
density distribution of the gas is approximately Gaussian (Paper I), and
the planar distribution of the CSDD resembles a double Gaussian (Eq. 
[\ref{eq:Equation.For.Sigma.C}]), the line-of-sight integration through
this density distribution is approximately double Gaussian as well.  In
nearly edge-on geometry the two Gaussians merge (the butterfly `closes
its wings').  The observable width, convolved with the synthesized beam
($FWHM_{\rm b}$), is given by:

\negMEDskip
\begin{eqnarray}
&& \hspace*{-10mm} FWHM_{\rm obs}^2(R;V_{\rm chan}) \; = 
   FWHM_{\rm b}^2 \; + \nonumber \\
&& \hspace*{-2mm}
   \left[ FWHM_{\rm z}(R) \sin{i(R)} \right]^2 \; + \nonumber \\
&& \hspace*{-2mm}
   \left[ FWHM_{\rm p}(R;V_{\rm chan}) \cos{i(R)} \right]^2
   \; , \label{eq:Observable.Thickness}
\end{eqnarray}
\negMEDskip

\noindent Using this equation, a fit to the measured apparent widths in
many channels yields the intrinsic thickness and the inclination of the
gas layer.  The required values for $FWHM_{\rm p}(R;V_{\rm chan})$ are
determined from the CSDD maps.  In each channel map (and CSDD map) a
major axis position is chosen such that galactocentric radius $R$ is
sampled [via Eq.  (\ref{eq:Rproj})].  I solve Eq. 
(\ref{eq:Observable.Thickness}) using a non-linear least squares
technique\footnote{\label{voet-LSQ} The performance of a least squares
technique can be greatly improved by imposing physical limits on the
fitted parameters.  For example I restrict the inclination to lie in the
range 0\ad-90\ad, and the thickness of the gas layer to be positive.  A
standard LSQ routine (e.g.  MRQMIN, Press \etal 1990) can be easily
modified to include physical limits in the step where a new guess for
the solution is calculated: if the new guess exceeds a physical
boundary, then the new guess is placed halfway between the old guess and
the physical boundary.} which also yields the errors on $i$ and
$FWHM_{\rm z}$.  For the case of NGC 4244, typically 12 determinations
of $FWHM_{\rm obs}(R)$ are used.  The square of the difference between
the observed and model widths is minimized by varying the inclination
and gaseous width, where each data point is weighed by the square of the
inverse of its error.  The error on each data point is calculated by
adding (in quadrature) the error in the right-hand side of Eq. 
(\ref{eq:Observable.Thickness}), as calculated from the errors in $i$
and $FWHM_{\rm p}$, to the observational error.  Since the contribution
from the right-hand side depends on the exact value of the inclination,
several iterations are required. 

Previous workers (Sancisi \& Allen 1979; van der Kruit 1981; Rupen
1991a) corrected their edge channel flaring determinations for some of
these projection effects, but made no effort to model the emission in
lower velocity channels in detail.  However, as shown above, the
measured apparent widths in the lower velocity channels provide enough
additional information to simultaneous determine the flaring and
inclination.  Furthermore, since many azimuths (channels) are used, the
effects of local irregularities are reduced.

For less inclined systems, inclination and thickness can be determined
from each channel individually using

\negMEDskip
\begin{eqnarray}
&& \hspace*{-15mm} \cos{i(R;V_{\rm chan})} =
   \frac{\Delta d(R;V_{\rm chan})}{2 Y_{\rm C}(R;V_{\rm chan})}
   \; , \label{eq:eqn.For.Cos.I} \\
&& \hspace*{-15mm} {\rm and } \nonumber \\
&& \hspace*{-15mm}
FWHM_{\rm z}^2(R;V_{\rm chan}) \; = \; \frac{1}{\sin^2{i(R)}} \; \times 
   \nonumber \\
&& \hspace*{-10mm}
      \left( FWHM_{\rm obs}^2(R;V_{\rm chan}) - FWHM_{\rm b}^2
      \right. \; + \nonumber \\
&& \hspace*{-7mm}
      \left. 
      \left[ 2.35 W_{\rm C}(R;V_{\rm chan}) \cos{i(R)} \right]^2 \right)
   \; , 
   \label{eq:Solution.For.Thickness}
\end{eqnarray}
\negMEDskip

\noindent where $\Delta d$ is the projected separation of the two
``wings'' of the ``butterfly''.  From the inclination and thickness
values determined from the individual channels the weighted averages are
calculated, where the weights are the inverse of the square of the
errors.  The errors in the right hand sides of Eqs. 
(\ref{eq:eqn.For.Cos.I}) and (\ref{eq:Solution.For.Thickness}) include
the errors on $d$, $FWHM_{\rm obs}$, $Y_{\rm C}$, $W_{\rm C}$ and $i$.

Both the intrinsic thickness of the gas layer and the planar width of
the CSDD increase approximately linearly with radius: the former because
of the $1/R^2$ DM-halo density distribution (Paper I), the latter
because in the wedge-approximation of the CSDDs the path through this
wedge is proportional to $R$.  Simple analytic considerations (Olling \&
van Gorkom 1995\OvGnote; Olling 1995b, Chapt.  4) indicate that the
ratio of these two projected widths depends strongly on inclination and
only weakly on the properties of the galaxy.  For a typical channel, the
projected gas layer width exceeds the projected CSDD width for
inclinations larger than about 60 degrees.  This inclination cutoff is
confirmed using ``observations'' of simulated galaxies seen at various
inclinations (Fig.  \ref{fig:width.inclination.at.not90}; Olling 1995b,
Chapt.  4).  Thus it is expected that for galaxies with inclinations
exceeding 60\ad \ the flaring,

\noindent and hence the DM-halo flattening, can be determined.

%%%%\negSMLskip
\subsection{The Thickness of NGC 4244's Gas Layer}
\label{sec-thickness.of.NGC4244s.gas.layer}
%%%%\negSMLskip

For all major axis positions ($x$) the emission profile perpendicular to
the major axis are investigated.  Gaussian fits to these profiles yield
the peak and width (Fig.  \ref{fig:NGC4244.the.raw.widths}) for all
positions.  All the thickness measurements presented in this figure, as
as well as the odd channels in between, are used in the determination of
inclination and thickness of the gas layer.  The peak of the fitted
Gaussians at each radius is plotted in the two outer panels, the
apparent thickness in the inner panels.  A fit is only performed if
there are at least three independent points with a signal to noise ratio
$\ge$ 4 in the profile.  The SW-NE asymmetry in the surface density
profile (Fig.  \ref{fig:Sigma.gas.R}) at 5.5 kpc and 6-8 kpc can be seen
in many of the individual channels, indicating that the SW-NE asymmetry
of the radial surface density distribution as determined using the
Abel-inversion technique extends over a large range in azimuth. 

Because inclination and thickness are calculated rather differently in
the nearly edge-on case [Eq.  (\ref{eq:Observable.Thickness})] and the
not edge-on case [compare Eqs.  (\ref{eq:eqn.For.Cos.I}) and
(\ref{eq:Solution.For.Thickness})], stringent criteria are applied to
avoid spurious two-component fits.  A reliable two-component fit must
satisfy the following criteria: a) the amplitudes of the components may
not differ by more than 60\%, b) the components must have equal widths,
c) the separation between the components must exceed $MAX(FWHM_{\rm b},
0.77 \times FWHM_{\rm cmp})$, and d) its goodness of fit must exceed the
goodness of fit of the one-component fit.  Generally, the profiles are
well represented by a single Gaussian, so that, the thickness and
inclination are determined using Eq.  (\ref{eq:Observable.Thickness}). 
The results are presented, for the two sides of the galaxy, in Fig. 
\ref{fig:Thickness.and.Inclination}.  The radial variation of thickness
in the upper panel, the inclination in the lower panel.  The plot on the
left-hand side results if the inclination is a free parameter, while
fixing the inclination leads to the thickness determinations plotted on
the right.

Using the observational errors only, the reduced $\chi^2$ values of the
fits [to Eq.  (\ref{eq:Observable.Thickness})] are of the order of 3.5
(\pmt 1.5) for radii smaller than $\sim$ 8 kpc and increase sharply to
$\sim$ 13 (\pmt 7) beyond, indicating that either the model is bad or
that the errors have been underestimated.  In Appendix
\ref{sec-The.effects.of.kinematical.distortions} I show that
non-circular motions contribute to the error budget of Eq. 
(\ref{eq:Observable.Thickness}) as well.  The inclusion of errors
associated with non-circular motions ($\sim$ 2 \kms \ for $R\le 8.5$ kpc
and $\sim$ 8 \kms \ beyond) decreases the reduced $\chi^2$ to values
around one.  The onset of the warp around 8.5 kpc suggests that modest
non-circular motions are associated with the warp.  Other explanations
are possible however: for example the surface density distribution may
be strongly asymmetric in azimuth so that the \HI \ emission arises from
either near or far side rather than from both sides, or the warp may be
asymmetric like the warp of NGC 4565 (Rupen 1991b).

The beamsize corrected widths inferred from the high (10\asd1) and
intermediate (38\as) resolution data sets are generally equal to within
the errors, which indicates that the width measurements are not hampered
by resolution effects.  It is likely that the non-circular motions
inferred in the inner 5 kpc (\Sec
\ref{sec-rotcur.and.velocity.dispersion}, Fig. 
\ref{fig:Rotation.curves}), which might result from the ``spiral
structure'' as seen in the \HI surface density distribution (Fig. 
\ref{fig:Sigma.gas.R}), are not limited to the regions close to the
major axis.  Because the determination of the thickness of the gas layer
is less accurate in the presence of non-circular motions\footnote{In
fact, in Appendix \ref{sec-The.effects.of.kinematical.distortions} I
show [Eq.  (\ref{eq:apx.for.delta.FWHM_C})] that the errors in the
derived thickness are proportional to the magnitude of the non-circular
motions and to the inverse of the rotation speed, which is rising in the
inner 5 kpc.}, it is not advisable to use this region to draw inferences
about the mass-to-light ratio of the stellar disk.  The derivation of the
halo's flattening, which uses the flaring data beyond 8 kpc, is not
affected.

Another problem area may be the 7 - 8.5 kpc hump on the south-western
side, where the thickness of the gas layer is 70\% larger, and the
gaseous surface density distribution (Fig.  \ref{fig:Sigma.gas.R}) 50\%
lower than on the north-eastern side.  This region is adjacent to a
region where the stellar surface density is possibly depressed a factor
of ten (\Sec \ref{sec-optical.data} and Fig. 
\ref{fig:radial.light.profile}).  In the outer part of the optical disk
the self-gravity of the gas can contribute significantly to the vertical
force (see Paper I, Fig.  10).  In the case of NGC 4244, models which
include the gaseous self-gravity have gas layer widths 50\% smaller than
models which do not, between 6 and 10 kpc.  Therefore I plot the
north-eastern rather than the average values in the final flaring curve
(Fig.  \ref{fig:Thickness.of.the.gas.layer}) from from 6 to 8.5 kpc.

From these flaring measurements taken at different resolutions and
sensitivities I extract two flaring curves, one for the varying
inclination case, the other for the fixed inclination case (Fig. 
\ref{fig:Thickness.of.the.gas.layer}).  For radii between 6 and 10 kpc I
use the flaring data extracted from a 357-pc resolution data set.  The
high (177 pc) and intermediate (665 pc) resolution flaring results are
used for the regions smaller than six and larger than ten kpc,
respectively.  The ``errors'' are either the weighted error on the
average of the two sides, or if larger, half the difference between the
two values, so that the {\em systematic} differences are reflected by
the extent of the error bars.

%%%%\negSMLskip
%\subsubsection{Inclination effects}
\subsection{Inclination effects}
\label{sec-Inclination.effects}
%%%%\negSMLskip

Because the inferred short-to-long axis ratio of the halo is
proportional to the square of the true width of the gas layer (Paper I,
Paper III), the thickness of the gas layer has to be measured reliably. 
It must be stressed that the inclination is a critical parameter for
determining the intrinsic thickness of the gas layer.  This is so
because, at inclinations around 85\ad \ a large part of the apparent
width arises due to the projected separation of the CSDD wings (\Sec
\ref{sec-Outline.of.the.method}), which changes by 40\% between 85\ad
and 83\ad.  At these inclinations the DM-halo shape can be determined
reliably only if the inclination is accurately known (Paper III).

%%%%\negSMLskip
%\subsubsection{Non-circular Motions}
\subsection{Non-circular Motions}
\label{sec-Non-circular Motions}
%%%%\negSMLskip

The assumption of circular orbits is basic to the thickness
determination technique presented in this paper.  Closed orbits of
ellipticity $\epsilon_{\phi, \rm p}$ exist in elliptic logarithmic
potentials, for which the radial and tangential velocity can be found
analytically [e.g., Rix \& Zaritsky 1995, their Eqs.  (7) and (8)].  As
a result, the difference between the observable {\em recession}
velocities of an elliptical disk and a circular disk have a period of
$\pi/3$ and an amplitude of $\epsilon_{\phi, \rm p} V_{\rm rot}$.  With
$\theta_{\phi}$ the direction of the major axis of the potential, the
observable velocity difference between two points at ($R,\theta$) and
($R,-\theta$) equals zero and $\left( 2\epsilon_{\phi, \rm p} V_{\rm
rot} \sin{(2\theta) \sin(\theta + 2\theta_{\phi})} \right)$ for a
round disk and an elliptical disk, respectively.  For ellipticities as
reported by Rix \& Zaritsky (0.045$^{+0.03}_{-0.02}$), the effects of
elliptical orbits are smaller than 9 \kms, so that the shape of the CSDD
maps, and hence the determination of the flaring is not affected much. 
The recession velocity at the major axis equals $V_{\rm sys} \pm V_{\rm
rot}(R) \left\{ 1 + \epsilon_{\phi, \rm p}
\cos{(2\theta_{\phi})}\right\}$.  Given the predicted and measured upper
limits to $\epsilon_{\phi, \rm p}$, ($\leq 0.1$, Dubinski 1994; and
$\leq 0.075$, Rix \& Zaritsky 1995), the measured 15\% decrease in
rotation speed cannot be explained by a rapid radial variation in
ellipticity or location of major axis of the dark matter distribution.

%\input{conc_02}
%
% File: conc_02.tex
%
%%%%\negMEDskip
%\subsection{DISCUSSION}
\section{DISCUSSION}
\label{sec-Discussion}
%%%%\negMEDskip

The thickness of the gas layer has been measured for only a few
galaxies: The Milky Way (e.g., Kulkarni \etal 1982; Knapp 1987;
Wouterloot \etal 1990; Diplas \& Savage 1991; Merrifield 1992; Malhotra
1994, 1995), NGC 891 (Sancisi \& Allen 1979; van der Kruit 1981; Rupen
1981a), NGC 4565 (Rupen 1991a), and M31 (Braun 1991).  The rapid
variations (up and down by a factor of 2 - 4 within 500 pc) in the
thickness of the gas layer of NGC 891 (Rupen 1991a) and the 7-kpc hump
found in the flaring curve of NGC 4244 (Fig. 
\ref{fig:Thickness.of.the.gas.layer}) are probably caused by spiral arm
streaming motions which change the shape of the emitting regions of the
channels (see also Appendix
\ref{sec-The.Effects.of.Spiral.Arm.Streaming.Motions}).  Inside the
stellar disks, the thickness ($FWHM$) of the gas layer increases
gradually from $\sim$150 pc to $\sim$800 pc at the edge of the stellar
disk, similar to the flaring of NGC 4244 as determined in this paper. 

Inside the optical disk ($R \leq 10$ kpc), the radial variation of the
dispersion is quite similar to that observed in other, more face-on,
systems (van der Kruit \& Shostak 1982; Dickey \etal 1990; see Kamphuis
1993, Chap.  12, for a review).  Beyond the optical disk ($R > 10$ kpc),
$\disp{gas}{}$ increases somewhat (from 8.5 to 9.2 \kms), similar to the
dispersion increase (from 6 to 8 \kms) seen in NGC 628 (Kamphuis 1993,
Chap.  12, Fig.  1).  In the other two systems where measurements of the
dispersion beyond the optical disk exist, $\disp{gas}{}$ declines very
slowly, to 6 and 7 \kms \ (Dickey \etal 1990; Kamphuis 1993).  Thus, the
radial variation of velocity dispersion in the gaseous medium of NGC
4244 is not unusual. 

Rather than determining the best fit \MoverL{} and dark-halo's core
radius and central density I use the procedure outlined in Paper I to
find the range in stellar mass-to-light ratios which yield acceptable
fits to the observed rotation curve (the halo's flattening is not a
parameter in this procedure).  I then calculate the reduced $\chi^2$
values on a grid which covers this range in \MoverL{}.  Recall that for
given observed rotation speeds at 2.3\hR \ ($V_{\rm rot}(2.3\hR)$ and
8\hR \ ($V_{\rm rot}(8\hR))$ and a choice for the stellar \MoverL{} the
dark halo parameters are uniquely determined (Paper I).  However,
because measurement errors on the rotation curve, different values
values for $V_{\rm rot}(2.3\hR)$ and $V_{\rm rot}(8\hR)$ are acceptable. 
I adopted values which differ by less than 1-$\sigma$ from the adopted
rotation curve which yield the lowest $\chi^2$ values.  In Fig. 
\ref{fig:disk.halo.decomposition} I present the so found `best fit' and
the \pmt 1-$\sigma$ mass models.  The mass-to-light ratio is rather well
determined (\MoverL{B} = $3.71^{+2.6}_{-0.4}$).  The upper limit
coincides with the maximum-disk value (van Albada \& Sancisi 1976), for
which the peak of the rotation curve due to the stellar disk equals the
observed peak rotation speed.

The shape of the vertical distribution of the stars determines the
inferred mass-to-light ratio of the stellar disk (van der Kruit 1988). 
Using a sech$^2$ vertical distribution, the inferred mass-to-light ratio
is twice larger than when using a vertically exponential distribution
(for edge-on galaxies).  As the \MoverL{B} value found here (6.3) for
the maximum disk case is about twice as large as typical for a
maximum-disk Scd galaxy (Broeils 1992, Chapt.  10) the vertical
distribution may be exponential rather than sech$^2$.  Additional
information (e.g., K-band imaging) regarding the vertical structure of
the stellar disk is required to determine \MoverL{} more accurately. 

Beyond the optical disk, the dynamical mass-to-light ratio increases to
a value of 14.9 \Msun/\Lsun \ at the last measured point (16 kpc), or
half that value for the vertically exponential distribution.  Neither
value is unusual (Broeils 1992, 1995).  Only a small amount of dark
matter is present: 1.4 times the luminous mass for the maximum disk
case, and 3.6 times the luminous mass for the half-maximum disk case
(independent of the vertical distribution).  As a result of the limited
extend of the measured rotation curve, the dark-halo parameters are ill
constrained: to within a factor 50 for the core radius and 900 for the
central density.

%%%%\negMEDskip
%\subsection{CONCLUSIONS}
\section{CONCLUSIONS}
\label{sec-Conclusions}
%%%%\negMEDskip

The rotation curve of the almost edge-on galaxy NGC 4244, determined
from high resolution \HI spectra, rises slowly to about 100 \kms at 5
kpc, remains flat till the edge of the optical disk at 10 kpc and
decreases in Keplerian manner till the last measured point at seven
optical scale lengths (14 kpc).  The mass-to-light ratio of the stellar
disk lies between 50\% and 100\% of the maximum-disk value, the core
radius and central density of the dark halo could not be determined.

I developed a new technique to determine inclination and flaring for
large \HI rich galaxies seen at inclinations larger than about 60\ad. 
Kinematic information is used to disentangle flaring and warping.  The
thickness of NGC 4244's gas layer increases gradually from $\sim$ 400 pc
at 2 optical scale lengths to $\sim$1.5 kpc at the last measured point.

\acknowledgments

  I like to thank Jacqueline van Gorkom for many valuable suggestions to
improve this paper, for guidance in image processing, for advice in
bicycle racing, for support and encouragement, and for discussing many
interesting aspects of extra galactic astronomy.  I would like to thank
NRAO and the Kapteyn Astronomical Institute, for their hospitality as
well as for their excellent software support.  I thank NRAO for the
generous allocation of observing time for this project.  It was a great
pleasure to discuss galactic structure and many aspects of \HI observations
of galactic disks with Michael Rupen.  I thank Juan Uson for discussing
various aspects of calibrating VLA data and providing us with the
wonderful UV-editing tasks UVLIN and UVMLN.  Piet van der Kruit helped
me a lot by providing the B-band photometry.  I also thank John Hibbard,
Mike Merrifield and the referee for an unbiased reading of a previous
version of this paper.  This research has made use of the NASA/IPAC
Extragalactic Database (NED) which is operated by the Jet Propulsion
Laboratory, CALTECH, under contract with the National Aeronautics and
Space Administration.  This work was supported in part through an NSF
grant (AST-90-23254 to J.  van Gorkom) to Columbia University.

% Appendices
%\input{app_02a}
%
% Appendices, File: app_02a
%

\subsection*{APPENDICES}

\appendix

\section{DETERMINING THE ROTATION CURVE AND VELOCITY DISPERSION OF AN
EDGE-ON GALAXY}
\label{sec-Theoretical.spectra}

In a manner similar to the tangent-point method employed for the Milky
Way (e.g., Burton \& Gordon 1978; Celnik \etal 1979), the gaseous
velocity dispersion and the circular velocity can be derived from the
extreme velocity tails of spectra taken on the major-axis of an edge-on
galaxy.  In this section I investigate how gradients (along the
line-of-sight) in surface density, rotation speed and velocity
dispersion affect this derivation.  Small gradients occur naturally in
the azimuthally symmetric galaxy models considered in this paper, while
strong gradients can occur near spiral arms.  Gradients perpendicular to
the line-of-sight, beam smearing, are also addressed. 

For a galaxy, in which all gas moves at exactly the circular velocity
($V_{\rm c}$), gas on the major axis moves at the local rotation
velocity.  Other positions have lower projected velocities, $V_{\rm p}$
[Eq.  (\ref{eq:Rproj})].  In reality, the velocity dispersion of the
gas broadens the sharp cutoff at $V_{\rm c}(R)$ and moves the peak of
maximum emission closer to the systemic velocity.  The spectrum,
$I(x,V,d;S)$, at a point\footnote{The coordinate system is defined in
\Sec \ref{sec-coordinate.system}.} $(x,d)$ on the sky depends on the
surface density distribution of the gas, the distribution perpendicular
to the plane, the gaseous velocity dispersion, the rotation curve, and
the channel response function $S(V';V,\Delta V)$, defined by Eq. 
(\ref{eq:channel.response.function}).  After summing the data along the
minor axis, the dependence on minor axis position and the vertical
distribution disappear so that the spectrum (or ``XV plot'') is given
by:

\negMEDskip
\begin{eqnarray}
&& \hspace*{-4mm}
   I(x,V;S) \; = \;  \frac{K}{\sqrt{2\pi}}
   \int dy \; \frac{\Sigma(y)}{\sigma(y)} \nonumber \\
&& \hspace*{-14mm}
   \int dV' \; S(V';V,\Delta V) \;
      e^{
      \left(-\frac{1}{2}
         [(V'-V_{\rm p}(y))/\sigma(y)]^2
      \right)
      } ,
      \label{eq:eqn.for.I.edge.on}
\end{eqnarray}
\negMEDskip

\noindent with $K$ a proportionality constant related to the choice of
units, $y = \sqrt{R^2 - x^2}$, and

\negMEDskip
\begin{eqnarray}
%y    &=& \sqrt{R^2 - x^2} \; \; , \\
\hspace*{-3mm}
V_{\rm p}(y) &=&   V_{\rm c}(R) \times \frac{x}{\sqrt{x^2 + y^2}}
   \nonumber \\
   \hspace*{-3mm}
   &\approx& V_{\rm c}(x) + \frac{y^2}{2x^2}( a x - V_{\rm c}(x))
           \; \; .
   \label{eq:eqn.for.V.y}
\end{eqnarray}
\negMEDskip

\noindent Here $V_{\rm c}(R) \approx V_{\rm c}(x) + a \; (R-x)$, and $a$
the gradient in the rotation curve (\kms /kpc).  A point somewhere along
the line of sight whose projected velocity is $\sigma$ \kms \ lower
than the rotation speed will hardly influence parts of the spectrum at
velocities larger than $V_{\rm c}(x)$.  Thus, the spatial integral in
Eq.  (\ref{eq:eqn.for.I.edge.on}) can be limited to a maximum, $\pm
y_{\rm max}$ so that only gas ``close'' to the major axis will
contribute significantly to the extreme velocity tails (EVTs).  Eq. 
(\ref{eq:eqn.for.V.y}) provides an estimate for $y_{\rm max}$:

\negMEDskip
\begin{eqnarray}
y_{\rm max} &\approx& x \;
   \sqrt{ - \;
      \frac{2\; \sigma(x)} {a x - V_{\rm c}(x) }
   }
   \; \; . \label{eq:eqn.for.epsilon.max}
\end{eqnarray}
\negMEDskip

\noindent For NGC 4244, typical values for $\epsilon_{\rm m}(=y_{\rm
max}/x)$ decrease from 0.85 at 1 kpc to 0.33 at 12 kpc.  The velocity
dispersion and the surface density can be expanded around $R=x$ as well:
$\sigma(y) \approx \sigma(x) \; 
\left( 1 + \frac{1}{2} \; b x \epsilon^2 \right)$, and 
$\Sigma(y) \approx \Sigma(x)+\frac{1}{2} \; c x \epsilon^2$,
where $b$ is the fractional gradient of the velocity dispersion
(kpc$^{-1}$), and $c$ the gradient in surface density (\MSpcsq/kpc).

For spectral line data obtained with the VLA which are not Hanning
smoothed (as is the case for the observations presented in this paper),
a channel of width $\Delta V$, centered at $V$, is sensitive to emission
at other velocities $V'$ as well.  The channel response function is
given by (Rots 1990):

\negMEDskip
\begin{eqnarray}
\hspace*{-7mm}
S(V';V,\Delta V) &=& 
   \frac{ \sin{\left( \pi (V'-V)/\Delta V \right)} }{ \pi (V'-V)/\Delta V }
   \; \; . \label{eq:channel.response.function}
\end{eqnarray}
\negMEDskip

\noindent The strong sidelobes of $S$ are reduced by more than an order
of magnitude when integrated over the width of a channel (the 1$^{st}$,
2$^{nd}$, 3$^{rd}$, and 4$^{th}$ channels have an effective sensitivity
of 100\%, +8.6\%,-1.9\%, and +0.83\%, respectively).  Limiting the
velocity range (to $\pm \frac{1}{2}\Delta V$) of the second integral in
Eq.  (\ref{eq:eqn.for.I.edge.on}) and taking $S$ to be unity over this
interval allows for an analytic approximation of $I(x,V;S)$, which can
then be used to gauge the importance of line-of-sight gradients of the
variables. 

Approximating the Gaussian term in Eq.  (\ref{eq:eqn.for.I.edge.on}) by
a parabola and integrating the spatial integral up to $\epsilon_{\rm
m}$, the spectrum can be written as:

\negMEDskip
\begin{eqnarray}
&& \hspace*{-8mm}
I(x,V) \; \approx \;  
   \left( \frac{ \epsilon_{\rm m} \; K \; \Delta V \; \Sigma }
               { \sqrt{2\pi} \sigma}
   \right) \; \;  e^{-\frac{1}{2}\left( \frac{W}{\sigma} \right)^2}
   \times \nonumber \\*[3mm]
&& \hspace*{-14mm}
   \left\{ 2 \; - \; \frac{\epsilon^2_{\rm m} W V_{\rm c}}{3\sigma^2} \; +
      \frac{x \epsilon^2_{\rm m}}{3} \; \times
   \right. \nonumber \\
&& \hspace*{-10mm}
   \left.
      \left[
         \left( \frac{a W}{\sigma^2}\right) \; + \;
         b \left( \frac{W^2}{\sigma^2} - 1 \right) \; + \; \frac{c}{\Sigma}
      \right]
   \right\}
   \; , \label{eq:approximation.for.spectra}
\end{eqnarray}
\negMEDskip

\noindent where all galaxian properties are evaluated at radius $R=x$,
and with $W = V-V_{\rm c}(x)$.  The multiplicative term in Eq. 
(\ref{eq:approximation.for.spectra}) is a Gaussian function of
dispersion $\sigma$.  The first term within the curly brackets results
from material on the major axis, the second term arises due to material
close to the major axis, while the last three terms originate due to
radial gradients in rotation speed, dispersion and surface density,
respectively.  Except in the inner 6 kpc, where beam smearing is
important as well, Eq.  (\ref{eq:approximation.for.spectra}) represents
the measured EVTs well.  A beam smearing correction could be obtained by
from an approximation of the convolution of the spectrum
(\ref{eq:approximation.for.spectra}) with the beam.  However, as this
leads to quite ugly equations, another approach is followed.  As we have
seen above, the EVTs of an XV plot contain information about the
intrinsic rotation curve and velocity dispersion.  A Gaussian fit to
such an EVT yield a value for the position and width of this
Gaussian\footnote{In practice multiple components are fitted to the full
spectrum such that the EVT is dominated by one component.  Because of
the additional components (more degrees of freedom) the errors are
overestimated.  The AIPS tasks SLICE and SLFIT were used to perform the
Gaussian fits.  However, the version of these tasks which are part of
the standard AIPS release re-sample every slice to a 512 grid, as a
result of which the calculated errors are meaningless.  I therefore
wrote special versions of SLICE and SLFIT which do not re-sample the
data.}.  Line-of-sight and beam smearing effects cause these values to
deviate from the intrinsic rotation speed and velocity dispersion.  The
magnitude of these deviations can be empirically found by comparing the
values obtained from fits to a simulated\footnote{Using the procedure
outlined in Appendix \ref{sec-VLA.simulate}.} XV plot with the
parameters used in the calculation of the simulated XV plot.  I find:

\negMEDskip
\begin{eqnarray}
&& \hspace*{-16mm}
\disp{gas}{} \approx
   \frac{\sigma_{\rm XV;10}}{\left( 0.9 + {\rm exp}[-R/4] \right)}
   \label{eq:Beam.Smearing.sigma.10}  \\*[3mm]
&& \hspace*{-16mm}
\disp{gas}{} \approx
   \frac{ \sigma_{\rm XV;38,55} }
      {\left( 0.9 \; {\rm exp}[-2.5] +  {\rm exp}[(R-10)/17]\right)} \; ,
   \label{eq:Beam.Smearing.sigma.38.55}
\end{eqnarray}
\negMEDskip

\noindent for $(R \leq 10 {\rm kpc})$ and $(R > 10 {\rm kpc})$
respectively.  The subscripts ``10'', ``38'', and ``55'' refer to the
beamsize of the data set for which the correction is appropriate.  The
``XV'' subscript refers to the value as fitted to the extreme velocity
envelope of the XV plot.  Note that the true gaseous velocity dispersion
is always smaller than the dispersion of the extreme velocity envelope. 
As these corrections depend on the input parameters (e.g., beam smearing
is insignificant for a flat rotation curve), the input parameters for
the calculation of the simulated XV plot should be close to the
intrinsic rotation and dispersion curves: this problem is solved by
iterating several times.  An estimate of $\sigma_{\rm XV}$ can also be
obtained by counting the number ($N_{\rm 0,P}$) of velocity channels
between the zero and peak emission level of the EVT.  This estimate is
calibrated by generating Gaussian distributions with a range of
dispersions centered at various locations within a velocity channel:
($\sigma_{\rm XV} = (2.4 N_{\rm 0,P} - 0.6 \pm 1) \times \frac{\Delta
V_{\rm chan}}{5.2}$ \kms).  Since $N_{\rm 0,P}$ as determined from the
data is generally equal to the value determined from the best-fit
simulated XV plot, I conclude that the velocity dispersion (and rotation
curve, see below) as determined from the Gaussian-fit method is
reliable. 

Because these relations depend on beam size, channel width, surface
density distribution, and the rotation curve, these relations are
specific to the observations of NGC 4244 presented in this paper.  The
errors, $3\sigma$, about 20\% and 5\% in the inner and outer part of the
galaxy, are largest in those regions with strong surface density
gradients.

The corrections required to determine the rotation curve from
the XV-plots are given by:

\negMEDskip
\begin{eqnarray}
&& \hspace*{-14mm}
V_{\rm rot;10\; \; \; }(R) \; \approx \; V_{\rm XV;10} \; \; \; +
   \nonumber \\
&& \hspace*{-10mm}
   \left( 1.3\times{\rm exp}[-R/4.0] + 0.35 \right) \times \sigma_{\rm XV;10}
   \label{eq:Beam.Smearing.Vrot.10}   \\*[3mm]
&& \hspace*{-14mm}
V_{\rm rot;38,55}(R) \; \approx \; V_{XV;38,55} + 
   \nonumber \\
&& \hspace*{-10mm}
   \left( 1.0\times{\rm exp}[-R/4.0] + 0.35 \right) \times \sigma_{\rm XV;38,55}
   . \label{eq:Beam.Smearing.Vrot}  
\end{eqnarray}
\negMEDskip

\noindent In regions where the surface density changes rapidly, the
smearing correction is least reliable: differences between input and
output rotation curves of order 3 - 5 \kms \ occur in the inner 5 kpc of
the simulated XV plots.

These correction are then applied to the fitted rotation and dispersion
values obtained from the data.  The resulting dispersion and rotation
curves are shown in Figs.  \ref{fig:gaseous.velocity.dispersion} and
\ref{fig:Rotation.curves}, respectively.  At some locations, the data
EVTs are very different from the simulated EVTs, while adjacent
positions have similar EVTs: these locations are likely to be affected
by local irregularities and have been discarded. 

This procedure does not work perfectly, not even for the azimuthally
symmetric model galaxy used here.  Both the XV-derived rotation curve and
the XV-derived dispersion curves (after applying the corrections
specified above) deviate systematically from the input curves.  Between
1 and 2 kpc and between 4.5 and 5 kpc the XV-derived rotation curve
falls $\sim$8 \kms \ ($\sim$12\%) and $\sim$ 3 \kms \ ($\sim$3\%) below
the input curve.  These negative differences can be clearly attributed
to the strong density enhancements at slightly larger radii, which are
thus seen at lower projected velocities: on the ``inside'' of strong \HI
density enhancements the XV-velocities are decreased and the true
rotation speeds are underestimated.  Too large a rotation speed (3 \kms)
is derived in the region where the \HI distribution strongly peaks (Fig. 
\ref{fig:Sigma.gas.R}), between 3.5 and 4.5 kpc.  This \HI peak behaves
like an $\delta$ function on the major axis, in which case the
XV-velocity directly corresponds to the rotation speed: applying the
dispersion correction [Eq.  (\ref{eq:Beam.Smearing.Vrot.10})] thus leads
to an overestimate of the rotation speed.  In determining the rotation
curve of NGC 4244 (\S \ref{sec-rotcur.and.velocity.dispersion}) I have
not tried to compensate for the surface density effects in detail since
these \HI density peaks may well be spiral arms, in which case the
derived rotation speeds may include some spiral arm streaming motions as
well.  Instead I have followed the upper envelope of both sides and in
the regions where the model calculations indicate that too low a
rotation speed is inferred from the XV-plots I increased the adopted
rotation curve such that is a smooth junction of the adjacent regions. 
Here the errors are accordingly increased.

Because beam smearing effects are not important beyond 6 kpc, Eq. 
(\ref{eq:approximation.for.spectra}) can be used to investigate the
dependence of the shape of the EVTs on gradients in $V_{\rm c}(R)$,
$\Sigma(R)$, and $\sigma(R)$.  Since for a given range in $y$ (or
$\epsilon_{\rm m}$) only a small range in $R$ is sampled, gradients in
the rotation curve are unimportant.  Gradients along the line of sight
in surface density and dispersion cause changes in $V_{\rm XV}$ and
$\sigma_{\rm XV}$.  Using Eqn.  (\ref{eq:approximation.for.spectra}) and
determining $V_{\rm XV}$ and $\sigma_{\rm XV}$ as a function of $b$ and
$c$, I find the following offsets (in \kms):

\negMEDskip
\begin{eqnarray}
&& \hspace*{-16mm}
\Delta V_{\rm XV} \approx (-3 \pm 0.7) \; b \; \; , 
   \label{eq:eqn.for.Delta.Vxv} \\
&& \hspace*{-16mm}
\Delta \sigma_{\rm XV} \approx
   (1.8 \pm 0.8) b +  (0.84 \pm 0.12)\left(\frac{c}{\Sigma}\right),
   \label{eq:eqn.for.Delta.sxv} 
\end{eqnarray}
\negMEDskip

\noindent for $( c \leq 0 )$, while the $c$-dependence in
(\ref{eq:eqn.for.Delta.sxv}) is absent for $c > 0$.  Where the
$\Delta$-symbol represents the difference between the ``no gradients''
and the ``strong gradients'' XV-parameters.  Beyond the optical disk,
typical values for $c/\Sigma$ and $b$ are -0.5 and 0.05, respectively. 
The smearing corrections [Eqs.  (\ref{eq:Beam.Smearing.sigma.10})
through (\ref{eq:Beam.Smearing.Vrot})] take these gradients into
account.  From Eq.  (\ref{eq:eqn.for.Delta.sxv}) it follows that in
order to {\em over} estimate \disp{XV}{} (and hence \disp{}{}) large
{\em positive} line of sight gradients in the dispersion are required. 
With \disp{XV}{} of the order of 10 \kms, a 10\% over estimate is
possible if the dispersion along the line of sight increases to 15 \kms
\ over a distance of 1 kpc.  If such a strong \disp{}{}-gradient would
exist in the radial direction as well, it would certainly be noticed. 
Very large velocity dispersion gradients have to exist to affect $V_{\rm
XV}$ significantly.  I conclude that the rotation and velocity
dispersion curves can be determined reliably from XV-plots of edge-on
galaxies.

%%%%\negSMLskip
%\subsection{APPENDIX-B: SIMULATING VLA SPECTRAL LINE OBSERVATIONS}
\section{SIMULATING VLA SPECTRAL \\
LINE OBSERVATIONS}
\label{sec-VLA.simulate}
%%%%\negSMLskip

In order to facilitate the interpretation of \HI spectral line data I
developed software which simulates VLA \HI observations of galaxies. 
The accuracy of the determination of rotation and velocity dispersion
curves from XV plots, the beam-smearing corrections, the
thickness-and-inclination determination technique, can all be gauged by
comparing the ``observations'' of such a simulated galaxy with the input
parameters.  Input parameters to this program (VLASIM) are: the
observing geometry, $\Sigma(R)$, $V_{\rm rot}(R)$, and \disp{gas}{}. 
The vertical density distribution of the gas layer is calculated
employing a physical model for the potential of the galaxy (the
multi-component method, Paper I) which yields the gaseous volume
density.  The rotation speed is assumed\footnote{For several disk-halo
models, rotation curves at different z-heights were calculated, all
showing that the rotation speed does not vary significantly ($\leq$ 5
\%) in the region where gas is found ($\abs{z} \leq 3 FWHM_{\rm
z,gas}$).} to be constant with height above the plane.  VLASIM then
integrates the amount of hydrogen along lines of sight from the observer
through the galaxy, and distributes this amount over the appropriate
channels [using the channel sensitivity function Eq. 
(\ref{eq:channel.response.function})].  When the line of sight
integrations are done, the spectral line cube is smoothed to the
requested resolution followed by the addition of realistic noise (if so
desired).

%\input{app_02b}
%
% File: app_02b.tex
%

%%%%\negSMLskip
%\subsection{APPENDIX-C: THE EFFECTS OF KINEMATICAL DISTORTIONS}
\section{THE EFFECTS OF KINEMATICAL DISTORTIONS}
\label{sec-The.effects.of.kinematical.distortions}
%%%%\negSMLskip

As discussed in \Sec \ref{sec-the.thickness.of.the.gas.layer}, knowledge
of the rotation curve allows us to determine the intrinsic thickness of
the gas layer.  However, the technique is sensitive to deviations from
the assumed rotation curve.  Spiral arm streaming motions of several
tens of percent of the rotation speed (several tens \kms) are not
uncommon in spiral galaxies (e.g., Visser 1980; Vogel \etal 1988;
Boulanger \& Viallefond, 1992; Tilanus \& Allen, 1993), so that locally
the velocity deviates from the circular velocity.  When a line of sight
is along a spiral arm, there will be a net shift in the radial velocity
(type 1), when the line of sight crosses a spiral arm, the radial
velocity will not be affected but the dispersion will be
larger\footnote{I thank Ron Allen for pointing this out.} (type 2).  In
this section I determine which parts of the galaxy are most susceptible
to non-circular motions and present a weighting scheme which minimizes
the effects of non-circular motions: a summary is presented in the next
section (\ref{sec-The.effects.of.kinematical.distortions.summary}).

The galactocentric coordinates [($x,y,z$) or \\ $(R,\theta,z)$] are
related to the celestial coordinates [(x,d) centered on the galaxy] in
the following way:

\negMEDskip
\begin{eqnarray}
x &=& R \cos{\theta} \; ,
   \label{eq:eqn.for.x} \\
d &=& R  \cos{i} \sin{\theta} + z \sin{i} \; .
   \label{eq:eqn.for.d}
\end{eqnarray}
\negMEDskip

\noindent where the position angle of the major axis was taken east-west
(90\ad) and where $z$ is the height above the plane.  If gas located at
$(R,\theta,0)$ has an excess line of sight velocity of

\negMEDskip
\begin{eqnarray}
&& \hspace*{-14mm}
\frac{\Delta V}{\sin{i}} = 
   \left[ V_{\rm rot}(R_0) \cos{\theta_0} - 
          V_{\rm rot}(R) \cos{(\theta_0 + \Delta \theta})
   \right]
   \nonumber
\end{eqnarray}
\negMEDskip

\noindent with respect to a position $(R_0,\theta_0,0)$, this position
will be ``seen'' in the same channel as $(R_0,\theta_0,0)$ is. 
Similarly, a channel $\Delta V$ \kms \ wide ``sees'' gas in an area
delimited by $\theta_0$ and $\theta$.  Expanding $\cos{(\theta_0 +\Delta
\theta)}$ to second order around $\theta_0$, and assuming that rotation
curve around $R_0$ is constant, I approximate the above equation and
solve for $\Delta \theta$. 

\negMEDskip
\begin{eqnarray}
&& \hspace*{-16mm}
   \frac{\Delta V \csc{i}}{ V_{\rm rot}(R_0)} \approx
   \left[
      \sin{\theta_0} \; \Delta \theta + 
      \frac{1}{2}\cos{\theta_0} (\Delta \theta)^2
   \right], 
   \label{eq:eqn.for.Delta.V} \\*[3mm]
&& \hspace*{-8mm} {\rm and} \nonumber \\
&& \hspace*{-16mm}
\frac{\Delta \theta}{\tan{\theta_0}} =
   \left[ -1 +
      \sqrt{ 1 +  
         \frac{2 \Delta V \csc{i}}{V_{\rm rot}(R_0)}
         \frac{\cos{\theta_0}}{\sin^2{\theta_0}}
      } \;
   \right] .
   \label{eq:eqn.for.Delta.theta}
\end{eqnarray}
\negMEDskip

\noindent The requirement that the two points [$(R_0,\theta_0,0)$ and
$(R,\theta,0)$] have the same major axis distance leads [via Eq. 
(\ref{eq:eqn.for.x})] to the following relation for $\Delta R \;(\equiv
R-R_0)$:

\negMEDskip
\begin{eqnarray}
\frac{\Delta R}{R_0}  &\approx& 
   \left(
     \frac{\Delta \theta}{1-\Delta \theta \; \tan{\theta_0} }
   \right) \; \tan{\theta_0}
   \label{eq:eqn.for.Delta.R}
\end{eqnarray}
\negMEDskip

\noindent In the lower panel of Figure
\ref{fig:The.effects.of.kinematical.distortions} I present the
relations, both practically independent of inclination, for $\Delta
\theta$ and $\Delta R$ graphically for the case where $\Delta V=10$ \kms
\ and $V_{\rm rot}(R_0)=100$ \kms.  Although this galaxy model is
extremely simple, the opening angles as determined for a more realistic
model behave similarly: the high velocity channels (small $\theta_0$)
are wider than the low velocity channels.  For $\theta_0 \leq 50\ad$ the
fractional range in galactocentric radius is small so that the
approximate relation for $\Delta V$ (\ref{eq:eqn.for.Delta.V}) is
reasonable, whether or not the rotation curve is constant around $R_0$. 

If the positions $(R_0,\theta_0,z_0)$ and $(R,\theta,0)$ project on the
same position on sky, and if due to an excess motion they also have the
same velocity, then their positions in the observer's spectral-line cube
will be the same.  The equivalent vertical height of this streaming
motion (or, equivalently, of a channel $\Delta V$ \kms \ wide) equals
$z_0$.  Using (\ref{eq:eqn.for.d}), (\ref{eq:eqn.for.Delta.R}) and
expanding sin($\theta$) around $\theta_0$ I find:

\negMEDskip
\begin{eqnarray} 
z_0 &\approx& R_0 \; 
   \left( \frac{\Delta \theta}{\cos{\theta_0} \tan{i}} \right) \; \; .
   \label{eq:theta.equivalent.to.zheight}
\end{eqnarray}
\negMEDskip

\noindent I plotted the projection of $z_0$ ($\equiv z_0 \sin{i})$ in
the second to bottom panel of Figure
\ref{fig:The.effects.of.kinematical.distortions}.  The curves are
labeled with the corresponding galactocentric radii.  For $i=85\deg \;$
its value is small compared to the projected vertical thickness of the
gas layer, but its importance relative to the projected vertical
thickness of the gas layer grows rapidly with decreasing inclination
($\propto \tan^{-1}{i}$). 

The systematic errors introduced by streaming motions are more easily
understood when considering the corresponding planar width of the CSDD, 
$\Delta Y(\Delta V) \approx R_0 \; \left( \frac{\Delta
\theta}{\cos{\theta_0} } \right)$, which can be approximated as:

\negMEDskip
\begin{eqnarray}
&& \hspace*{-12mm}
   \frac{\Delta Y(\Delta V) }{R_0}
   \approx
   \left[ \cos{\theta_0}\; \left( 
      1.81 {\cal V} \; \sin^{1.2}{\theta_0} \; + 
   \right. \right.
   \nonumber \\*[3mm]
   && \hspace*{+10mm}
   \left. \left.
      \sqrt{ {\cal V} \cos{\theta_0} } \; \right)
   \right]^{-1} ,
   \label{eq:eqn.for.Delta.Y}
\end{eqnarray}
\negMEDskip

\noindent with ${\cal V} \equiv \left[ (V_{\rm rot}(R_0) \; \sin{i})/
(2\Delta V) \right]$.  In the region close to the major axis, kinematical
distortions have an effect proportional to the square root of the
fractional velocity deviation (2$^{nd}$ term), in the lower velocity
channels the proportionality is approximately linear (1$^{st}$ term). 
Thus, the channels close to the major and minor axes have long
pathlengths through the galactic disk and are hence easily affected by
kinematical or surface brightness irregularities (Fig. 
\ref{fig:The.effects.of.kinematical.distortions}, third panel from
below).  The edge-channel itself is even wider as the two sides (at $\pm
\theta_0$) merge together. 

%\section{The Effects of Streaming Motions}
\label{sec-The.Effects.of.Spiral.Arm.Streaming.Motions}

Streaming motions of type 1 cause the centers of the CSDD to shift along
the line of sight, typically by $\pm \Delta Y$ [see also Eq. 
(\ref{eq:apx.for.delta.Y_C})].  In the almost edge-on situation, the two
sides of the CSDD, as projected on the sky, are overlapping so that the
projected planar width [Eq.  (\ref{eq:Equation.For.FWHMp})] and its
error are dominated by the $Y_{\rm C}$ term.  Visser (1980) presents
model velocity fields for M 81 which include spiral arm streaming
motions (his Fig.  10).  The equivalent CSDD map results after
deprojection of the velocity field (see also \Sec
\ref{sec-Outline.of.the.method}).  From these velocity fields it is
clear that the centers of the CSDDs shift substantially due to spiral
arm streaming motions: outwards on the outside of the arms and inwards
on the inside of the arms.  The strong kinks follow the spiral arm and
would produce sudden changes in the separation of the near and far sides
of the CSDDs and in the width thereof, over a large range in azimuth. 
If streaming motions are present, but the resulting CSDD-kinks are not
taken into account in the analysis, the inferred widths may show strong
gradients when crossing a spiral arm.  Such may be the reason for the
large thickness gradients seen in NGC 891 (Rupen 1991a) and NGC 4244
(the 7-kpc hump, \Sec \ref{sec-thickness.of.NGC4244s.gas.layer}).

At lower inclinations, when the two sides of the CSDD are separated,
streaming motions of type 2 affect the width determination [Eq. 
(\ref{eq:apx.for.delta.FWHM_C})], while the errors on $Y$ (type 1
streaming motions) hamper the determination [via Eq. 
(\ref{eq:eqn.For.Cos.I})] of the inclination.  This is illustrated in
the top panel of Figure
\ref{fig:The.effects.of.kinematical.distortions}, where I plot the
inclination error resulting from a velocity shift of 10 \kms:

\negMEDskip
\begin{eqnarray}
\Delta i &\approx& 
   \frac{\cos{i}}{\sin^2{i}} \;
   \frac{\Delta \theta}{\sin{\theta_0}\cos{\theta_0}}
   \label{eq:eqn.for.Delta.i}  
   \; \; . \label{eq:equation.for.delta.i}
\end{eqnarray}
\negMEDskip

\noindent Thus, small streaming motions occurring close to either the minor or
the major axis can cause large errors in the inferred inclination. 
Furthermore, the inclination errors are strongly inclination dependent:
$\Delta i(i=65\ad) \approx 6 \Delta i(i=85\ad)$.

\negSMLskip
\subsection{Summary}
\label{sec-The.effects.of.kinematical.distortions.summary}
\negSMLskip

I will use the above determined systematic errors estimates in the
thickness \& inclination determination process (\Sec
\ref{sec-Outline.of.the.method}).  The effect of spiral arm streaming
motions is to either shift the center of the CSDD along the line of
sight or to increase the apparent velocity dispersion, the former having
more severe effects.  To our advantage, the expected systematic errors
scale with the width of a CSDD wing (calculated for a given gaseous
velocity dispersion \disp{gas}{}),

\negMEDskip
\begin{eqnarray}
&& \hspace*{-14mm}
\delta Y_{\rm C} \approx
   \frac{\Delta Y(\Delta V)}{\Delta Y(\disp{gas}{})} \times 
   FWHM_{\rm C}(\disp{gas}{})
   \label{eq:apx.for.delta.Y_C}  \; \; , \\
{\rm and} && \nonumber \\
&& \hspace*{-18mm}
\delta FWHM_{\rm C} \approx
   \left( 
   \sqrt{ 1 + \left( \frac{\Delta Y(\Delta V)}{\Delta Y(\disp{gas}{})} 
              \right)^2 } - 1 
   \right) \times \nonumber \\
&& \hspace*{-18mm}
   FWHM_{\rm C}(\disp{gas}{})
   \label{eq:apx.for.delta.FWHM_C} \; ,
\end{eqnarray}
\negMEDskip

\noindent are easily calculated from the CSDD model maps and Eq. 
(\ref{eq:eqn.for.Delta.Y}).  These systematic errors increase towards
lower inclinations and are proportional to $(\sin{i} \;
\sin{\theta}\cos{\theta})^{-1}$: the regions around both principal axes
are most easily affected by streaming motions.

%%%%\negSMLskip
%\subsection{APPENDIX-D: OPTICAL DEPTH EFFECTS}
\section{OPTICAL DEPTH EFFECTS}
\label{sec-Optical.Depth.Effects}
%%%%\negSMLskip

In the Milky Way the optical depth in the \HI line is substantial within
a few degrees from the galactic plane (e.g., Dickey \etal 1983). 
Typically, the total emission along the line of sight is under estimated
by 15\%.  Similar values are found for edge-on external systems (Haynes
\& Giovanelli 1984).  We thus expect similar values for NGC 4244.  Braun
\& Walterbos (1992) estimate that the cold interstellar medium, which is
mostly responsible for the absorption (see below), has a volume filling
factor of about 1 - 8\%, for the Milky Way and M31, respectively.  This
cold material forms a layer two to three times thinner than the warm \HI
(e.g., Kulkarni \& Heiles 1988; Braun \& Walterbos 1992).  For NGC 4244,
we thus expect that emission profiles perpendicular to the major axis
might be affected by absorption on the midplane, resulting in larger
apparent widths.  However, from the peak surface brightness and apparent
width measurements (Fig.  \ref{fig:NGC4244.the.raw.widths}) such effects
are not obvious: profiles with a high peak column density do not seem to
be broader.  A more detailed, albeit idealized, calculation of the
optical depth is given below. 

The optical depth in the \HI line for a uniform medium of length $l$ and
number density $n$ is given by Braun \& Walterbos.  Their Eq.  (7)
reads:

\negMEDskip
\begin{eqnarray}
\tau &=& 170 \frac{ n \; l}{T_{\rm s,100} \; \Delta V} \; \; ,
   \label{eq:eqn.for.tau}
\end{eqnarray}
\negMEDskip

\noindent with $T_{\rm s,100}$ the spin temperature in units of 100 K,
and $\Delta V$ the width of the channel in \kms.  In \Sec
\ref{sec-Inferred.properties} the surface density, the rotation curve
and the velocity dispersion were derived assuming low optical depth. 
Those derived quantities can be used to evaluate the quantities in Eq. 
(\ref{eq:eqn.for.tau}) and check the validity of the assumption of low
optical depth.  The CSDD maps provide the \HI surface density for which
the projected velocity equals the observing velocity ($\pm \frac{1}{2}
\Delta V$).  Investigating these CSDD channel maps (Fig. 
\ref{fig:CSDDs}) we find typical peak column densities of 0.5 - 1
\MSpcsq \ per channel for galactocentric radii smaller than 7 kpc, and
much less beyond.  In edge-on geometry, the surface density distribution
along the line of sight is a double Gaussian [Eq. 
(\ref{eq:Equation.For.Sigma.C})] with a typical width ($FWHM_{\rm p}$)
for each component of 3 \pmt 1 kpc.  The \HI volume number density is
given by: $n = 5.5 \; 10^{-3} \left( \Sigma_1 / FWHM_{\rm z,1} \right)$
cm$^{-3}$, where the units for surface density and thickness are \MSpcsq
\ and kpc.  The optical depth can then be rewritten as:

\negMEDskip
\begin{eqnarray}
\tau &=& 0.156
   \frac{ \Sigma_1 \; FWHM_{\rm p,3}}
        { FWHM_{\rm z,1} \; T_{\rm s,100}} \; \; ,
   \label{eq:eqn.for.tau.scale}
\end{eqnarray}
\negMEDskip

\noindent where a channel width of 6.2 \kms \ was used.  Typical values
for the spin temperature of the cold component of the interstellar
medium range from 100 - 175 K (for the Milky Way and M31, respectively;
Braun \& Walterbos 1992).  Smaller values are found in individual clouds
(Spitzer 1978, p.  48, and references therein).  Taking the model widths
as representative for the thickness of the gas layer ($\sim$ 250 pc,
Fig.  \ref{fig:Thickness.of.the.gas.layer}), and assuming that the cold
medium fills the entire volume we find $\tau \sim 0.33$ (30\%
attenuation).  If the cold medium is clumpy with a low filling factor,
as in the Milky Way, we might expect 3 clouds per kpc with an optical
depth $\geq$ 1 (Kulkarni \& Heiles 1988).  In an exactly edge-on
geometry this density of clouds would correspond to a large optical
depth.  However, at an inclination of 85\ad the path through the galaxy
($\sim$ 3 kpc) corresponds to a change in $z$ of $\sim 260$ pc (about
three times the thickness of the cold gas layer), so that the number of
absorbing clouds per line of sight is reduced.  Because the pathlengths
$FWHM_{\rm p}$ are comparable for all velocity channels, the optical
depth effects are comparable in all channels (in the inner 7 kpc).

Beyond 7 kpc the surface densities are much smaller so that optical
depth effects are not important in the derivation of the shape of the DM
halo (Paper III).

%\input{figc_02a}
%
% File: figc_02a.tex
%
%\setcounter{figure}{0}

\clearpage
\normalsize

\onecolumn

\sfRO
\begin{figure}
%%%%\plotone{olling_2.fig01.ps}
\caption[The Optical image \& total \protect\HI map]{
\label{fig:IIIaJ.image}
This image of NGC 4244, taken on a IIIaJ-plate, was kindly made
available by Piet van der Kruit.  See van der Kruit \& Searle (1981a)
for details.  The weak dustlane, barely visible on a TV monitor, could
not be reproduced in this greyscale image.  The contours ( [ 2 4 8 16 ]
x 6 10$^{20}$ cm$^{-2}$) represent the total \protect\HI column
densities at a resolution of 10\as $\!$x10\as (175x175 pc).  A greyscale
representation of the total \protect\HI map is presented in Fig. 
\protect\ref{fig:NGC4244.total.HI}. 
}
\end{figure}
\tfRO

%%%\twocolumn

\sfRO
\begin{figure}
%%%%\plotfiddle{olling_2.fig02.ps}{8cm}{0}{100}{100}{0}{0}
\caption[The radial light profile]{
\label{fig:radial.light.profile}
In this figure I present four radial light profiles, each again composed
of three profiles: SW (dotted line), NE (dashed line), and average (full
line).  For clarity, the profiles have been offset by a constant C. 
From top to bottom the profiles are derived from: a) all light (C=0.69),
b) all light within 420 pc from the major axis (C=-2.81), c) all light
more than 420 pc ``above'' (east of) the midplane (C=-6.31), and d) all
light more than 420 pc ``below'' (west of) the midplane (C=-9.81). 
Extinction is not important in the determination of the radial scale
length (\hR) because all profiles are consistent with the same radially
exponential distribution, with \hR=2.0 kpc (the filled circles). 
}
\end{figure}
\tfRO

%%%%\clearpage

\sfRO
\begin{figure}
%%%%\plotfiddle{olling_2.fig03.ps}{5cm}{0}{75}{75}{-10}{-330}
\caption[The total \protect\HI distribution]{
\label{fig:NGC4244.total.HI}
Two representations of the total \protect\HI distribution are shown.  A
low resolution (38\as $\!$x38\as, 665 pc) contour plot is superimposed
on a high resolution (10\as $\!$x10\as, 175 pc) grey scale image.  The
image has been rotated counterclockwise by 48\deg (NE is upp, SW is
down).  The units for the contours (0.1 0.4 0.8 1.6 3.2 6.4 12.8) and
the greyscales are 10$^{20}$ cm$^{-2}$.  A tickmark along the vertical
axis, 1\am, corresponds to roughly 1 kpc.  The slight warp sets in
around 9 arcmin, the edge of the optical disk.  Clear features in the
high resolution image are the the high surface brightness peaks seen on
both sides located approximately 6 arcmin from the center.  Notice an
approximately 2 kpc diameter region of low column densities, extending
from about +3 to +5 kpc.  The edges of this feature seem to extend to
larger height above the plane.  This feature is also clearly visible in
the channel maps (Fig.  \protect\ref{fig:channel.maps}). 
}
\end{figure}
\tfRO

%%%%\onecolumn

\sfRO
\begin{figure}
%%%%\plotone{olling_2.fig04.ps}
\caption[The Global Profile]{
\label{fig:global.profile}
The Global Profile, or the total \protect\HI flux as a function of
velocity, determined from the 55\as $\!$x55\as resolution data set.  The
systemic velocity is the mean of the 10\%, 20\% and 50\% points, 244.4
\pmt 0.4 \kms.  A small asymmetry between the approaching and receding
sides can be seen.  Typical uncertainties in the flux equal 0.025
Jy/channel. 
}
\end{figure}
\tfRO

%%%%\clearpage

\sfRO
\begin{figure}
%%%%\plotfiddle{olling_2.fig05a.ps}{14.9cm}{0}{106}{85}{-298}{-130}
\caption[An overlay of the high and low resolution channel maps]{
\label{fig:channel.maps} 
A contour representation of the channel maps at 38\as $\!$x38\as (665
pc) resolution superposed on a greyscale image of the channel maps at
28\asd3x10\asd1, (495x177 pc) resolution.  The greyscale levels
correspond to column densities of [1 1.5 3 6 9] $10^{20}\;$ cm$^{-2}$
(the lowest level is $\sim 2.6 \sigma$), the line contours to [-0.07
0.07 0.1 0.2 0.4 0.8 ] 10$^{20}$ cm$^{-2}$ (the lowest contour is $\sim
2.5 \sigma$).  The channels of the approaching and receding sides are
combined into one frame.  The value printed in the upper left corner
indicates the velocity offset from the systemic velocity (i.e.,
$\abs{V_{\rm chan}-V_{\rm sys}} \; / \; \Delta V_{\rm chan}$).  In order
to visualize the vertical structure better, the horizontal scale is
expanded.  The orientation is as in Fig. 
\protect\ref{fig:NGC4244.total.HI}, the distances are relative to the
pointing center listed in Tab. 
\protect\ref{tab:Table.N4244.obs.parameters},
}
\end{figure}
\tfRO

\sfRO
\begin{figure}
%%%%\plotfiddle{olling_2.fig05b.ps}{14.5cm}{0}{102}{90}{-298}{-130}
\figurenum{\protect{\ref{fig:channel.maps}b}}
\caption[More channel maps]{
The channels which show emission on the receding as well as on the
approaching side are shown individually here.  The
same coding of greyscales and line contours is used as in Fig. 
\protect\ref{fig:channel.maps}a .
}
\end{figure}
\tfRO

\sfRO
\begin{figure}
%%%%\plotfiddle{olling_2.fig06.ps}{15.5cm}{0}{100}{100}{0}{0}
\caption[The radial \protect\HI surface density distribution]{
\label{fig:Sigma.gas.R}
The gaseous surface density calculated using eqn
(\protect\ref{eq:Sigma.gas.R}, see \Sec
\protect\ref{sec-HI.distribution}), on a linear (lower panel) and
logarithmic scale (upper panel).  The full line, the dotted line, and
the dashed line represent the average, the south-western, and the
north-eastern surface density profiles.  Like in the Milky Way (Knapp
\etal 1978; Blitz, private communication), no sharp cutoff is observed
in the \protect \HI distribution. 
}
\end{figure}
\tfRO

\sfRO
\begin{figure}
%%%%\plotone{olling_2.fig07.ps}
\caption[The major-axis velocity (XV) plot]{
\label{fig:XV-plots}
The major-axis velocity (XV) plot of NGC 4244.  The lowest five contours
are derived from channel maps at low resolution (55\as), the highest
five contours from 10\asd1 resolution channel maps.  All emission within
2.5 kpc from the major axis was summed.  The two sides are remarkably
similar, both in kinematics and extent (to about 15 kpc or seven optical
scale lengths).  Only between +3 and +5 arcmin, the north-eastern side
shows a marked dip in the XV-plot.  This area is readily identified with
a lack of gas on the major axis in the channels 18 through 20 (see
Figure \protect\ref{fig:channel.maps}).  Note that this hole does not
``punch through'' the whole disk.  The derived rotation curve is
indicated by the filled circles. 
}
\end{figure}
\tfRO

\sfRO
\begin{figure}
%%%%\plotfiddle{olling_2.fig08.ps}{15cm}{0}{100}{100}{-320}{-200}
\caption[The radial variation of the gaseous velocity dispersion]{
\label{fig:gaseous.velocity.dispersion}
The beam-smearing corrected [using Eqs. 
(\protect\ref{eq:Beam.Smearing.sigma.10}) and
(\protect\ref{eq:Beam.Smearing.sigma.38.55})] tangential velocity
dispersion as derived from the high (10\asd1x10\asd1), intermediate
(38\as $\!$x38\as) and low (55\as $\!$x55\as) XV-plots (bottom, middle
and top panel respectively).  The south-western, north-eastern, and
adopted average curves are indicated by the dotted line, the dotted full
line and the filled circles respectively.  The measurements are
consistent with a constant velocity dispersion of roughly 8.5 km/s.  The
fat horizontal bars in the lower right corners indicate the size of the
beam. 
}
\end{figure}
\tfRO

\sfRO
\begin{figure}
%%%%\plotfiddle{olling_2.fig09.ps}{15cm}{0}{100}{100}{-320}{-200}
\caption[The Rotation Curve]{
\label{fig:Rotation.curves}
The beam-smearing and velocity dispersion corrected [using Eqs. 
(\protect\ref{eq:Beam.Smearing.Vrot.10}) and
(\protect\ref{eq:Beam.Smearing.Vrot})] rotation curve for the two sides
of the galaxy.  For an explanation of the symbols, see Fig. 
\protect\ref{fig:gaseous.velocity.dispersion}. 
}
\end{figure}
\tfRO

\sfRO
\begin{figure}
%%%%\plotfiddle{olling_2.fig10.ps}{12cm}{0}{90}{90}{-275}{-150}
\caption[Some Channel Surface Density Distribution (CSDD) maps]{
\label{fig:CSDDs}
This figure shows a representative set of CSDDs.  The channel numbers
are shown in the top right corner of each panel.  As the CSDDs are
symmetric with respect to the major axis (the horizontal axis), only
one quadrant is displayed.  The vertical axis of each plot represents
the distance along the line of sight (LOS axis).  The contours represent
the effective column density a given channel is sensitive to.  In this
coordinate system, the observer would be located at $(0,-\infty)$.  We
clearly recognize the wedge-like structure valid for the simple model
galaxy, however, the wedges have become wiggly and curve back towards 
the major axis (as a result of the declining rotation curve).  For
channels close to the extreme velocity channel (-19), the gaseous
velocity dispersion broadens the wedges significantly, and
fills up the gap between the near and far side. 
}
\end{figure}
\tfRO

%\input{figc_02b}
%
% File : figc_02b.tex
%

\sfRO
\begin{figure}
%%%%\plotfiddle{olling_2.fig11.ps}{15cm}{0}{100}{100}{-305}{-200}
\caption[Determining the thickness if $i\ne90\ad$]{
\label{fig:width.inclination.at.not90}
The results of simulated VLA \protect\HI spectral line observation of a
model galaxy like NGC 4244, inclined by 72\ad.  The simulated spectral
line cube is processed in the same manner as the observations of NGC
4244.  The recovered inclinations are shown in the lower panel, while
the recovered widths are presented in the top panel.  The filled circles
represent the input model widths.  No noise was added to the simulated
spectral line cube. 
}
\end{figure}
\tfRO

\sfRO
\begin{figure}
%%%%\plotfiddle{olling_2.fig12.ps}{13.5cm}{0}{95}{80}{-305}{-140}
\caption[The apparent widths of the channel maps]{
\label{fig:NGC4244.the.raw.widths}
In this figure the results of the Gaussian fits to the channel maps are
presented.  The middle two panels show the measured widths ($FWHM$, beam
size corrected, in units of kpc), the outer two panels show the fitted
peaks (\MSpcsq, beam size corrected).  The results obtained from the
(10\asd1 x 28\asd3, 177x495 pc) channel maps are represented by the open
squares, the fits to the intermediate resolution (38\as, 665 pc) maps by
the full line (with error bars).  At few positions two component fits
could be made, these are indicated by filled symbols (filled triangles
for the intermediate resolution data).  The two left-hand panels are
determined from the south-western side, the two right-hand panels from
the north-eastern side. 
}
\end{figure}
\tfRO

\sfRO
\begin{figure}
%%%%\plotfiddle{olling_2.fig13a.ps}{15cm}{0}{100}{100}{-305}{-210}
\caption[The flaring determined from the 10\asd1 resolution data]{
\label{fig:Thickness.and.Inclination}
Here I present the thickness (upper panel) and inclination (lower panel)
derived from the fitted widths (at 10\asd1 resolution) presented in Fig. 
\protect\ref{fig:NGC4244.the.raw.widths} using the procedure described
in \Sec \protect\ref{sec-Outline.of.the.method}.  Both sides
(south-western part on the left) of the galaxy are shown.  The fat bar
in the lower left corner of the top panel represents the size of the
beam, horizontally and vertically for the major axis and minor axis,
respectively.  For the panels on the left hand side I allowed the
inclination to vary, the panels on the right hand side are calculated
with a fixed inclination of 84.5\deg.  The filled circles represent the
average of both sides.  The weighting scheme which takes potential
errors due to streaming motions into account was used (see Appendix
\protect\ref{sec-The.effects.of.kinematical.distortions}). 
}
\end{figure}
\tfRO

\sfRO
\begin{figure}
%%%%\plotfiddle{olling_2.fig13b.ps}{15cm}{0}{100}{100}{-305}{-210}
\figurenum{\protect{\ref{fig:Thickness.and.Inclination}b}}
\caption[The flaring determined from the 38\as resolution data]{
Thickness and inclination as derived from the 38\as x 38\as resolution
channel maps.
}
\end{figure}
\tfRO

\sfRO
\begin{figure}
\plotfiddle{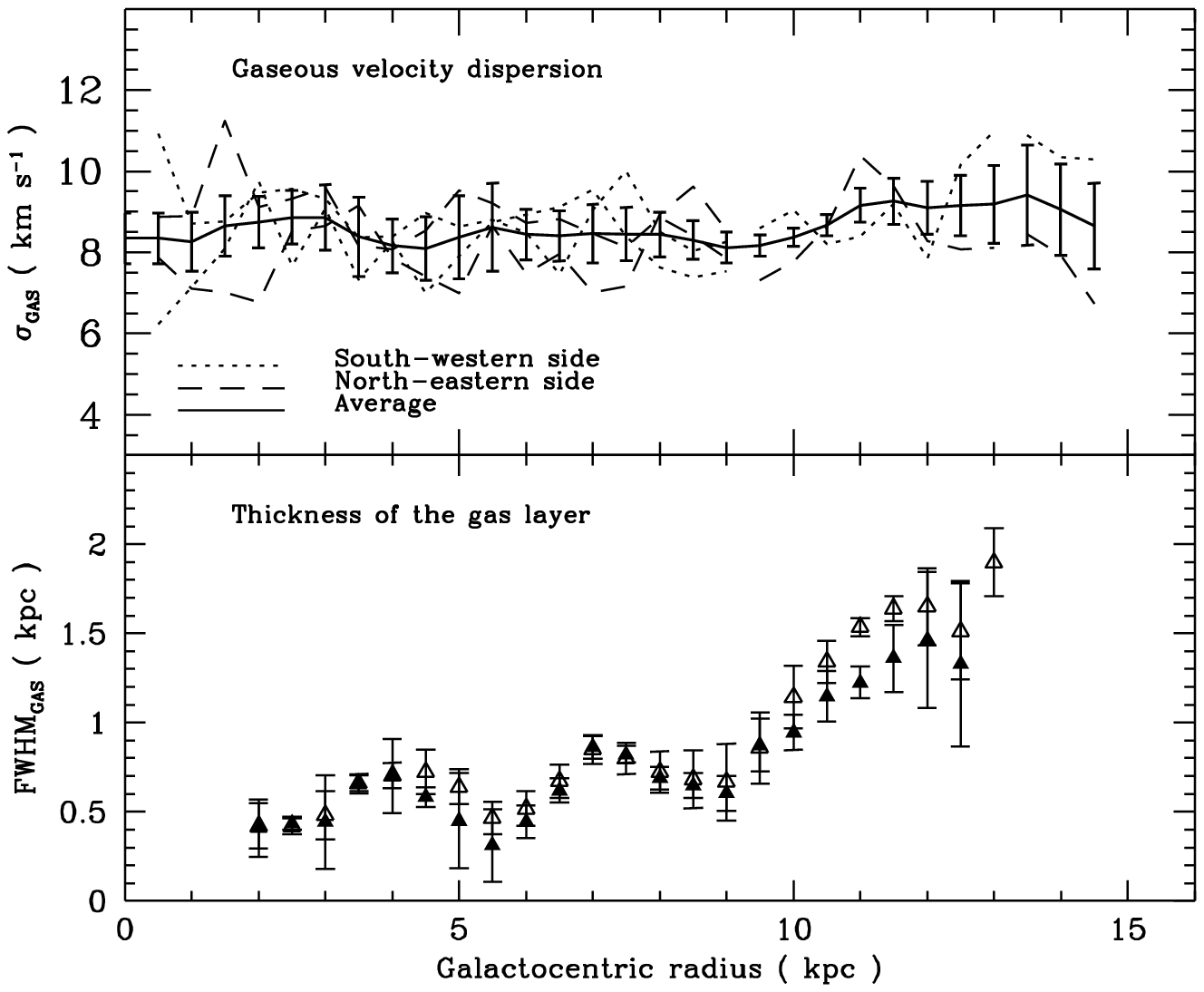}{15cm}{0}{100}{100}{-305}{-210}
\caption[The composite flaring curve]{
\label{fig:Thickness.of.the.gas.layer}
The lower panel shows the measured widths (open and filled triangles
with error bars, for the fixed and varying inclination case,
respectively).  The observed flaring curve is the average of the
north-eastern and south-western sides, except in the 6.5 - 9 kpc region,
where the north-eastern flaring data are used (see \Sec
\protect\ref{sec-rotcur.and.velocity.dispersion}) The 7 kpc hump is
probably caused by spiral arm streaming motions (Appendix
\protect\ref{sec-The.Effects.of.Spiral.Arm.Streaming.Motions}).  The top
panel displays the velocity dispersion measurements (full line with
error bars).  The dotted lines are determined from the south-western
side, the dashed lines from the north-eastern side (10\asd1 and 38\as,
for $R \leq 9$ kpc; 38\as \ for $R \in (9,13]$ kpc, and 55\as \ beyond). 
The adopted dispersion curve (full line with error bars) is the weighted
average of these individual curves. 
}
\end{figure}
\tfRO

\sfRO
\begin{figure}
\plotfiddle{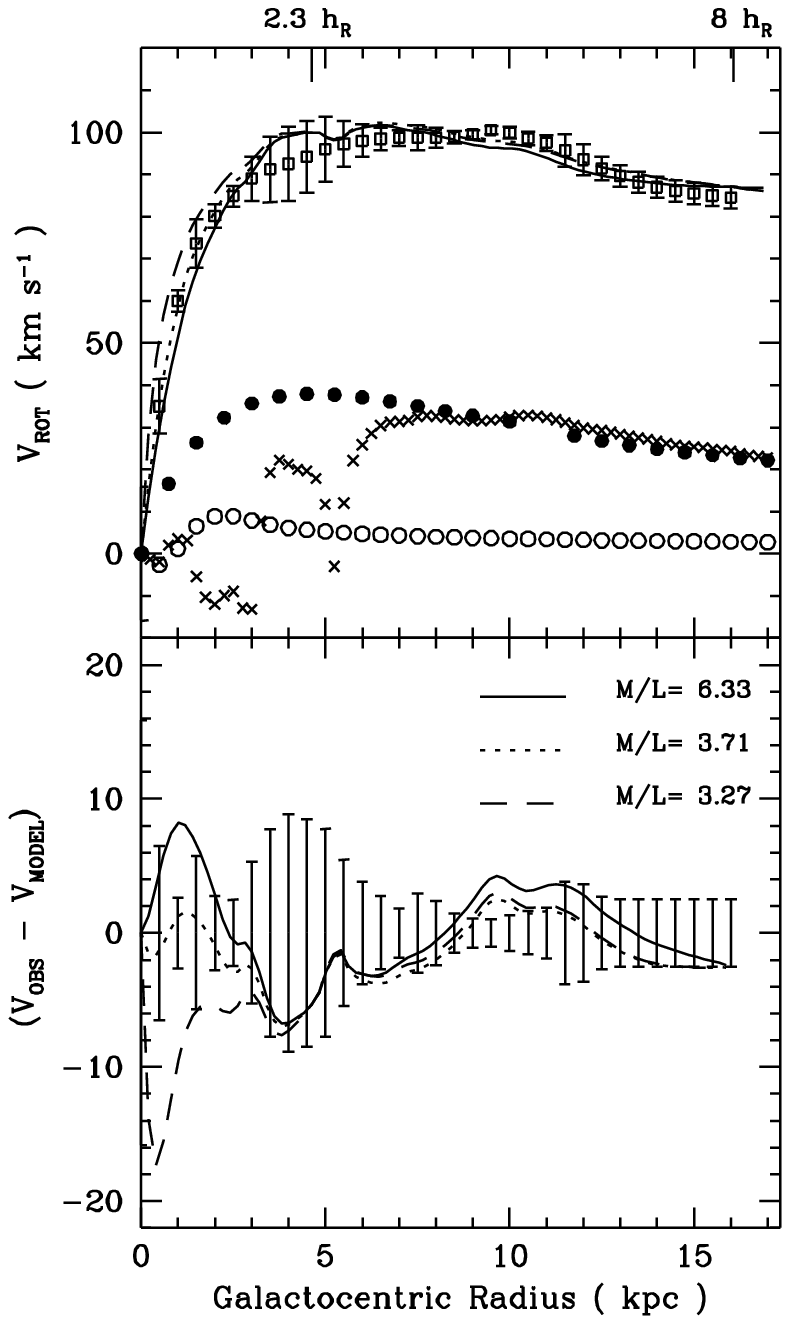}{11.5cm}{0}{100}{100}{-315}{-250}   
\caption[The disk-halo decomposition]{
\label{fig:disk.halo.decomposition}
The top panel shows disk-halo decompositions of the observed rotation
curve (squares with error bars) into components due to the gaseous disks
(\protect\HI,crosses; H$_2$, open circles), a component due to the
stellar disk (vertically sech$^2$ and radially exponential till $R_{\rm
max}$=10 kpc, \MoverL{B}=1; filled circles) and a DM component.  The
individual contributing DM components are not shown.  The data points
are independent out to 13 kpc, beyond the resolution equals 1 kpc.  The
rotation values at 2.3\hR \ and 8\hR \ are differ slightly (100 and 87
\kms) from the adopted rotation curve (Fig. 
\protect\ref{fig:Rotation.curves}) such as to minimize the residuals
(lower panel), see text.  The three disk-halo decompositions
($[\MoverL{B}=6.33, \Rc=13 \; {\rm kpc}, \rho_{\rm h,0}=1.43 \;
$m\Msun/pc$^3$] full line, $[\MoverL{B}=3.71, \Rc=0.65, \rho_{\rm
h,0}=233$] dotted line, $[\MoverL{B}=3.27, \Rc=0.28, \rho_{\rm
h,0}=1300$] dashed line) represent the best fit (\MoverL{B}=3.71,
$\chi^2$=0.864, 32 degrees of freedom) and those models with a reduced 
$\chi^2$ value 1.0 larger. For the disk-halo decompositions presented here, 
the $f_{\rm m}$, $f_8$, and $\beta_8$ parameters (as defined in Paper I) 
have the values 0.7094, 0.6033, and 0.861 respectively. 
}
\end{figure}
\tfRO

% \gamma = .973  .745 .699

\sfRO
\begin{figure}
%%%%\plotfiddle{olling_2.fig16.ps}{16cm}{0}{100}{77}{-305}{-200}
\caption[The effects of non-circular motions]{
\label{fig:The.effects.of.kinematical.distortions}
This figure illustrates the possible effects of streaming motions on the
apparent widths (see Appendix
\protect\ref{sec-The.effects.of.kinematical.distortions}).  All curves
presented scale roughly linearly with the ratio of streaming motion and
circular speed.  For CSDDs centered on $\theta_0$, I plot the angular
width (the lowest panel, crosses) and the normalized range in radius
(filled circles) which corresponds to a streaming motion of magnitude
$\Delta V$ \protect\kms.  The corresponding apparent width on the sky is
plotted in the one but lowest panel.  In the second panel from the top,
I plot the linear distance, in the plane of the galaxy, that the angle
$\Delta \theta$ corresponds to.  Close to the major and minor axes, long
pathlengths reflect a strong sensitivity to small streaming motions. 
The inclination errors resulting from such streaming motions are plotted
in the top panel. 
}
\end{figure}
\tfRO

\clearpage

\pagestyle{empty}

\sfRO
\begin{table}[hbt]
\caption{\label{tab:Table.N4244.parameters}
Optical parameters
}
\begin{tabular}{|l|l|l|}                                              \hline 
Distance                     &                          &  3.6   Mpc  \\
B-band radial scale length   & \hRb{,B}                 &  2.0   kpc  \\
B-band vertical scale 
   height exponential        & \ze(=$\frac{1}{2} z_0)$  &  0.21  kpc  \\
edge of optical disk         & $R_{\rm max,SW}$         &  10.0  kpc  \\
                             & $R_{\rm max,NE}$         &   8.0  kpc  \\
%limiting J magnitude        & &  27.5  mag \passq                    \\
face-on central 
   surface brightness        & $\mu_{0,\rm B}$          &  22.4 \pmt 
                                                           0.07 mag\passq \\
                             & L$_{\rm B}$(0)           &  74.6 \pmt 
                                                           5 L$_\odot \; 
                                                           {\rm pc}^{-2}$ \\
diameter of 25 mag\passq 
   isophote  & $D^i_{25}$    &  11    kpc                             \\
total B-band luminosity      & L$_{\rm tot,B}$          & 1.8 10$^9$ 
                                                          L$_\odot$  , 
                                                         $M_{\rm B}$ = -17.7
                                                                      \\ \hline
% plate scale                &  &  17.5 ps/\as, 1.05 kpc/\am, 0.952 \am/kpc \\
%L$_0$(B,exp)                &  & 149    L$_\odot$/pc$^2$                   \\
%$R_{25,\exp}$               &  &   6.3   kpc                               \\ 
%$\mu_{0,\rm exp}(B)$        &  &  21.6  mag/\assq                          \\ 
%L$_{\rm tot,B}$(exp)        &  &   3.7  10$^9$ L$_\odot$ , 
%                                   $M_{\rm B}$ = -18.4 \\
\end{tabular}
\tablenotetext{}{
Note: No correction for galactic extinction was required ($E(B-V)$=0;
Burstein \& Heiles 1984).  The subscript ``B'' refers to the value as
determined from B-band photometry.  The face-on central surface
brightness is calculated {\em assuming} a sech$^2(z/z_0)$ vertical light
distribution.  For a distribution which is exponential in the vertical
direction, L(0) is twice larger.  The absolute magnitude, the central
surface brightness, and the total luminosity, which are derived from the
central surface brightness and \hRb{,B} , change accordingly (to 21.6
mag\passq, 149 L$_\odot \; {\rm pc}^{-2}$ and $M_{\rm B}$ = -18.4,
respectively).  Note that $D^i_{25}$-values derived from the sech$^2$
and exponential luminosity density distributions (9.6 and 12.6 kpc)
bracket the statistically corrected RC2 value (de Vaucouleurs \etal
1976). 
}
\end{table}
\tfRO

\negBIGskip
\sfRO
\begin{table}[hbt]
\caption{\label{tab:Table.N4244.obs.parameters}
Instrumental parameters for the VLA observations of NGC 4244
}
\begin{tabular}{ll}  \hline \hline
Observing dates                  & \ 2 May 1989, B-array, \ 6 hours on source \\
                                 &  30 Nov 1989, D-array,  15 min  on source  \\
                                 &  29 Nov 1990, C-array,  45 min  on source  \\
                                 &  21 Aug 1992, D-array, \ 7 hours on source \\
Pointing center                  &  12\th  14\tm  59\tsd9              \\
                                 & 38\ad \ \ 5\am \ \ \ 6\asd0         \\
Velocity of band center          & 245  \kms, 
   heliocentric, optical definition                                    \\
Number of channels               & 63                                  \\
channel separation               &  5.2  \kms                          \\
Velocity resolution              &  6.2  \kms                          \\ \hline
map making characteristics       &  beam sizes: along and $\perp$ to the 
   major axis \\
\hspace*{5mm}  10\asd0 x 10\asd0 & B + C + D-array, uniformly
   weighted, tapered \\
\hspace*{5mm}  28\asd3 x 10\asd1 & B + C + D-array, uniformly
   weighted, tapered \\
%\hspace*{5mm}  20\asd5 x 20\asd5 & B + C + D-array, IMAGR, robust=0   \\
\hspace*{5mm}  38\asd0 x 38\asd0 & B + C + D-array, naturally weighted \\
\hspace*{5mm}  55\asd0 x 55\asd0 & 1992 D-array, naturally weighted    \\ \hline
rms noise in channel maps        &                                     \\
\hspace*{5mm}     10\asd0 x 10\asd0  & 
    $2.3\;\;$  mJy beam\rtp{-1} $\rightarrow$ 
    $131 \; \; \; 10^{18}$ cm\rtp{-2}                                  \\
\hspace*{5mm}     28\asd3 x 10\asd1  & 
    $1.9\;\;$  mJy beam\rtp{-1} $\rightarrow$ 
    $\;39 \; \; \; \; 10^{18}$ cm\rtp{-2}                              \\
%\hspace*{5mm}     20\asd5 x 20\asd5  & 
%    $0.63$   mJy beam\rtp{-1} $\rightarrow$
%    $\;\;8.9\; \;10^{18}$ cm\rtp{-2}                                  \\
\hspace*{5mm}     38\asd0 x 38\asd0  & 
    $0.73$   mJy beam\rtp{-1} $\rightarrow$ 
    $\;\;2.8\; \;10^{18}$ cm\rtp{-2}                                   \\
\hspace*{5mm}     55\asd0 x 55\asd0  & 
    $0.76$   mJy beam\rtp{-1} $\rightarrow$ 
    $\;\;1.4\; \;10^{18}$ cm\rtp{-2}                                   \\
    \hline \hline
\end{tabular}
\end{table}
\tfRO
\negMEDskip

\negMEDskip
\sfRO
\begin{table}[hbt]
\caption{\label{tab:Table.N4244.HI.parameters}
Parameters derived from the \protect\HI distribution
}
\begin{tabular}{|r|l|}                                \hline 
$\alpha_{{\rm adopted}}$(1950) &  12\th \    14\tm  \    59\tsd59     \\
$\delta_{{\rm adopted}}$(1950) &  38\ad \ \ \ 4\am  \ \  59\asd7      \\
%$l$                      & 154\add5719                         \\
%$b$                      & \ 77\add1584                        \\
Position angle           & -48\ad                               \\
Inclination              &  $\sim$ 84\add5                      \\
$R_{\HI}$                &  11.0   kpc                          \\ 
%Total Flux              & 440    \pmt \ 20 Jy \kms             \\
$V_{\rm sys}$, heliocentric & 244.4  \pmt 0.3 \kms              \\
M$_{\rm tot,\HI}$        &   (1.34 \pmt \ 0.07) 10$^9$ \Msun    \\
M$_{\rm tot,\HI}/L_{B}$  & 0.74 \Msun \Lsun$^{-1}$              \\
$W_{20}$                 & 180.4   \kms                         \\
$W_{50}$                 & 180.6   \kms                         \\ \hline
\end{tabular}
\tablenotetext{}{
Notes: The adopted center of mass is derived from the total \protect\HI
distribution (\Sec \protect\ref{sec-The.center.of.mass}).  The
\HI-radius, $R_{\HI}$, is the radius at which the \HI surface density
equals 1 \Msun \ pc\rtp{-2}.  $W_{50}$ and $W_{20}$ are the corrected widths
of the global profile at 50\% and 20\%, respectively, calculated
according to Botinelli \etal (1990).  . 
}
\end{table}
\tfRO

\end{document}